\documentclass{article}

\title{Human Social Cycling Spectrum}
\author{~\\ Wang Zhijian  and Yao Qinmei \\~\\
Experimental Social Science Laboratory, \\ Zhejiang University, Hangzhou 310058, China}
\date{\today}

\usepackage{graphicx}

 \usepackage{geometry}
\usepackage{indentfirst}
\usepackage{tabularx}     
\usepackage{amsmath}      
\usepackage{amsfonts}            
\usepackage{mathrsfs}            

\usepackage{color}    
\usepackage{graphicx}    
\usepackage{ulem}    
\newcommand{\tabincell}[2]{\begin{tabular}{@{}#1@{}}#2\end{tabular}}

\usepackage{hyperref}
\usepackage{multirow}
\usepackage{booktabs} 

\begin{document}

\maketitle

This paper investigates the reality and accuracy of
evolutionary game dynamics theory in human game behavior experiments.
In classical game theory, the central concept is Nash equilibrium,
which reality and accuracy has been well known
since the firstly illustration by the O'Neill game experiment in 1987.
In game dynamics theory, the central approach is dynamics equations,
however, its reality and accuracy is rare known,
especially in high dimensional games.
By develop a new approach, namely the eigencycles approach,
with the eigenvectors from the game dynamics equations,
we discover the high dimensional cycles in the same experiments.
We show that, the eigencycle approach can increase the accuracy
by an order of magnitude in data.
As the eigenvector is fundamental in dynamical systems theory
which has applications in natural, social, and virtual worlds,
the power of the eigencycles is expectedly.
Inspired by the eigencycles in O'Neill games, we suggest that,
the mathematical concept, namely 'invariant manifolds',
can be a candidate as the central concept for
the game dynamics theory,
like the fixed point concept (Nash equilibrium) in game statics theory.
%
%

~\\
~\\
Keywords: ~\\
behavior game experiment  ~\\
evolutionary game theory  ~\\
dynamics system theory  ~\\
eigenvector component~\\
eigencycle ~\\
invariant manifold ~\\

\newpage

\tableofcontents

\newpage
\section{Introduction}
\subsection{Research question}
We investigates the reality and accuracy of
evolutionary game dynamics theory \cite{2011Sandholm}\cite{dan2016book}
in high dimensional human game behavior experiments \cite{Behavioral2003}.

In game statics theory,
the mixed strategy Nash equilibrium
established around 1950 is the central concept.
Until 1987, a representative game experiment,
the O'Neill game\cite{1987Nonmetric},
provided the first illustration
that laboratory human strategy behavior
can be accurately captured by the central concept.
Table 1 shows the payoff matrix of
the experiments involving a long-run repeated,
discrete time, two-person, zero-sum game.
Since then, this game has been extensively
repeated in various experimental settings
 \cite{Binmore2001Minimax}\cite{Yoshitaka2013Minimax}
 and analysis \cite{Brown90}.
So far, the literature has focused on the long time strategy distributions
and the time dependence of individual behaviors,
but seldom explored social dynamics structure.
Here, the dynamics structure means
the geometrical pattern of
the evolutionary trajectories in the game state space.

Intuitively, two reasons hinder the study of the dynamics pattern.
First, the game has a high-dimension strategy space;
the means to handle the high dimensional dynamic pattern
remains an unresolved problem \cite{dan2020}.
Second, in a laboratory game experiment,
human strategy decision-making is highly stochastic,
and testing the dynamic pattern is a difficult task
\cite{Behavioral2003}\cite{1999What},
especially in a discrete time game \cite{dan2014}.
Past decade has seen the improvement on test out cyclic dynamical pattern
in some presentative games, like rock-paper-scissors \cite{dan2014,wang2014social}
and matching pennies-like $2\times 2$ game \cite{wang2014}.
But actually, all of these are two dimensional, not high dimensional.
In a real game, there always many players
and many strategy involve,
then the evolution trajectory will be in high dimensional state space.

To study high dimension game dynamics structure is not trivial.
As a cutting-edge question, not only in laboratory game experiments,
it appears in seeking the regularity in real game dynamics processes.
It is meaningful for the natural science and social science,
as well as in engineering and artificial intelligent visual world
\cite{dan2016book}\cite{2011Sandholm}\cite{dan2020}.
One can say that, in a real or visual world,
wherever the game theory can be applied,
during game playing, there always exists evolutionary processes;
Then in the processes which appearing as geometrical trajectories,
the regularity always a curious question.

\subsection{Background of eigencycle approach}
We thus develop an approach, the \textbf{eigencycle set}
to identify the high dimensional dynamics.
Our approach bases on
dynamical systems theory, a branch of mathematics.
As established before,
for the local dynamics near the equilibrium,
the linearization and
analytical solutions based on the
eigenvalues and eigenvectors
can be applied.
Suppose that an initial probability distribution can be expressed as
a linear combination of the eigenvectors $\xi$ as \cite{Lu5083},
\begin{equation}\label{eq:eigdeco1}
 {p}(0) =  a_1 {\xi}_1 + a_2 {\xi}_2 + ... + a_k {\xi}_k,
\end{equation}
wherein $\xi_i$ is associated eigenvector
of the eigenvalue  $\lambda_i$;
The probability will evolve in time according to
\begin{equation}\label{eq:eigdeco2}
{p}(t) = e^{\lambda_1t} a_1 {\xi}_1 + e^{\lambda_2 t} a_2 {\xi}_2
         + ... + e^{\lambda_k t} a_k {\xi}_k,
\end{equation}
where, the coefficients
$(a_i, i \in [1,k])$ and
$({\xi}_i, i \in [1,k])$ are independent of time $t$,
and the only time depending term is $e^{\lambda_i t}$.
Instated of the coefficients $a_i$  or the eigenvalue $\lambda_i$, we concern
the components of the eigenvectors.
Suppose a normalized eigenvector having $s$ components
and denoting as
${\xi}_i = (\eta_1, ..., \eta_m, ..., \eta_n, ... \eta_s)$,
we can present the $m$-th dimension evolution as
\begin{equation}\label{eq:eigdeco2}
{p_m}(t) = e^{\lambda_1 t} a_1 {\eta_m}_1 + e^{\lambda_2 t} a_2 {\eta_m}_2
         + ... + e^{\lambda_k t} a_k {\eta_m}_k.
\end{equation}
Similary, the $n$-th dimension evolution as
\begin{equation}\label{eq:eigdeco2}
{p_n}(t) = e^{\lambda_1 t} a_1 {\eta_n}_1 + e^{\lambda_2 t} a_2 {\eta_n}_2
         + ... + e^{\lambda_k t} a_k {\eta_n}_k.
\end{equation}
Obviously, $p_m(t)$ and $p_n(t)$ will form a trajectory in the two dimensional space
(hereafter denoted as $\Omega^{mn}$
which is a  subspace of the $s$ dimensional space).

Our starting point comes from the synchronization
between two components belong to a given eigenvector.
There exists two time invariant:
First, the phase angle difference of any two components
$\left(\arg(\eta_m)-\arg(\eta_n)\right)$ is invariant along time;
Second, the amplitude of each component $\eta_i$, $||\eta_i||$,
is invariant along time. Then we study
the  inference of two components in general conditions.
By which,
we develop a new theoretical expected observation,
called as eigencycle set,
to identify human dynamics behaviors in a high-dimensional game.

The main contributions are as follows:
(1) As illustrated in Figure \ref{fig:fine_structure},
we find the fine and hyperfine structures
in high-dimensional game dynamics pattern in a historic human game experiment data,
with the cycle measurement accuracy increased by an order of magnitude.
(2) We propose a tool, the eigencycle set analysis,
for general dynamics system analysis. We illustrate its validation
 in the O' Neill game data who firstly illustrated the accuracy of Nash equilibrium in 1987.   

Inspired by the finding in experiments as well as the approach
which deeply rooted in standard dynamics system theory,
we suggest that, the central concept of
game dynamics theory needs to be reconsidered,
which will be discussed later.

\subsection{Contents organization}
The rest of this paper is organized as follows:
In section 2,
we introduce the O'Neill game as the example,
and analytical calculate the eigen system
of the replicator dynamics equations; Then
we illustrate the  eigencycle set approach
and the theoretical results, 
from which the fine structure 
and hypefine structure are predicted;
In section 3,
we introduce the six human behavior game experiment data,
and report the experimental results of
the high dimensional cycle motion
projection on the two dimensional subspaces;
In section 4,
we validate the theoretical eigencycle set
 approach for bridging theory and experiments.
 In this section, the discovery of the fine structure and
 the significant evidence of the hyperfine structure are reported.
In section 5,
we summarize our contributions, the implication of our finding and the approach
and present the scope for future research. 
At the last, by numerical results comparisons, we suggest that, 
the 'invariant manifold' can be  
the potential central concept for game dynamics theory.

%
\section{Theoretical results on the eigencycle set}
\subsection{The eigenvector in the O'Neill game}
The O'Neill game is a zero-sum, 4$\times$4 game. Table \ref{tab:gamemodel}
shows the payoff matrix:
\begin{table}[h!]
\caption{The O'Neill zero-sum game matrix}
\begin{center}
\begin{tabular}{c|rrrr}
 \hline
&B1&B2&B3&B4\\
 \hline
A1&1&-1&-1&-1\\
 \hline
A2&-1&-1&1&1\\
 \hline
A3&-1&1&-1&1\\
 \hline
A4&-1&1&1&-1\\
 \hline
\end{tabular}
\end{center}
\label{tab:gamemodel}
\end{table}
To investigate dynamics behavior in a laboratory experiment game,
we use the replicator dynamics equation \cite{2011Sandholm}:
\begin{equation}\label{eq:repliequl} \Dot{x}_j=x_j*(U_j - {\overline{U_X}}),
\end{equation}

$x_j$ is the $j$th strategy player's probability
in the population with the $j$th strategy player included, and
$\Dot{x}_j$ is the evolution velocity of the probability;
 $U_j$ the payoff of the $j$th strategy player,
and $\bar{U}_X$ is the average payoff of
the population with the $j$th strategy player included.
The explicitly expression of these nonlinear differential equations
are shown in SI.
One by one, we assign strategy probabilities (A1, A2, A3, A4)
to $(x_1,x_2,x_3,x_4)$, respectively. Similarly,
we assign strategy probabilities (B1, B2, B3, B4)
to $(x_5,x_6,x_7,x_8)$, respectively.
Then, at any time, the system must be
an eight-dimension space, wherein
the Nash equilibrium is
$$x^*:=(x_1^*,x_2^*,...,x_8^*)=(2/5,1/5,1/5, 1/5,2/5,1/5,1/5,1/5).$$
This eight-dimension space has two concentrated
$$ x_1+x_2+x_3+x_4=1 ~ \cap ~ x_5 + x_6 + x_7 + x_8=1 \cap x_k \geq 0 ~ (k \in {1,2,...,8})$$
From the Jacobian matrix (or character matrix, or derivative matrix)
of the velocity vector field
$F:=(\Dot{x}_1,\Dot{x}_2,...,\Dot{x}_8)$ at Nash equilibrium $x^*$ \cite{2011Sandholm},
we can calculate the eigenvalues $\lambda$ and
their related normalized eigenvector $\xi$'s components
$(\eta_1,\eta_2,...\eta_8)$ explicitly.
Here, the components
$(\eta_1,\eta_2,...\eta_8)$ are one-by-one corresponding
to $(x_1,x_2,...,x_8)$.
The explicitly analytic results of
the eigenvalue $\lambda$ and eigenvector $\xi$
are shown in Table \ref{tab:eigenval_vec_cyc}.
From the sign of the eigenvalues
we can read off that the Nash eqilibrium (fixed point)
is linearly along the eigendirection of the real eigenvalues.
Details of the eigen system deduction are shown in appendix section \ref{app:derive_eig}.
\subsection{Eigencycle set}
\begin{flushleft}
\textbf{Definition}
\end{flushleft}
We call the circle constructed by the
two components $(\eta_m,\eta_n)$ within one normalized eigenvector
${\xi}_i = (\eta_1, ..., \eta_m, ..., \eta_n, ... \eta_s)$
the \textbf{eigencycle}, marked as $\sigma^{(mn)}$.  It is calculated as:
\begin{equation}\label{eq:the_ecyc}
 \sigma^{(mn)}=\pi \cdot ||\eta_m|| \cdot ||\eta_n||
               \cdot  \sin\big (\arg(\eta_m)-\arg(\eta_n)\big ),
\end{equation}
where $m$ and $n$ are the abscissa and the ordinate dimension
of the two dimension subspace, respectively.
$||\eta_m||$ and $\arg(\eta_m)$ indicate
the amplitude and the phase angle
of the $\eta_m$, respectively.
$\sigma^{(mn)}$ can determine the direction and amplitude of the eigencycle.
An alternative and equivalent presentations of the eigencycle value are
\begin{eqnarray}\label{eq:the_ecyc}
       \sigma^{(mn)}&=&\pi \cdot \big (\Re(\eta_m)\Im(\eta_n)
                                    - \Re(\eta_n)\Im(\eta_m)\big ),\\
                    &=&\pi \cdot \Re\big(\eta_m^\dag \eta_n \big)
\end{eqnarray}
wherein the $\Re(\eta_m)$ is the real part of
the complex number $\eta_m$ and
the $\Im(\eta_m)$ is the imaginary part;
the superscript $\dag$ indicates the conjugate
of the complex number.

\begin{figure}[h!]
\centering
\includegraphics[scale=0.7]{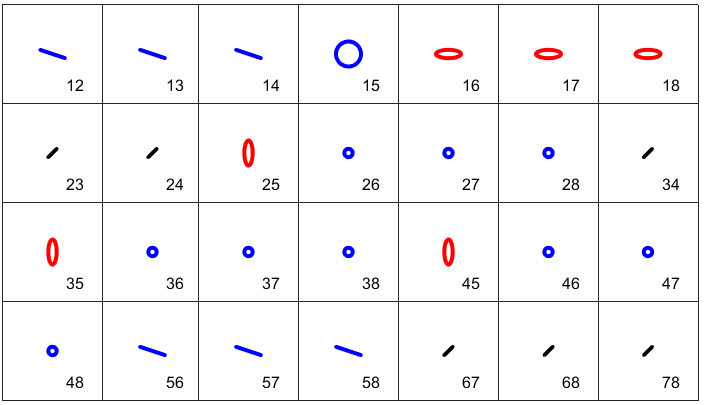}
    \caption{Geometric presentation of the eigencycle set. Subfig (mn) illustrate the 28 eigencycles patterns of the eigencycle set $(\sigma_{.8i})$; the value of eigenvecter component ($\eta_m$ and $\eta_n$) and cycles value $\Omega^{(mn)}$ is shown in Table \ref{tab:eigenval_vec_cyc}.
    The red (blue) color indicates the cyclic motion being counter-clockwise (clockwise),
    and its value (called as the signed area value or the angular momentum value), is positive (negative).}
\label{fig:Lissajous}
\end{figure}

\begin{flushleft}
\textbf{Eigencycle values of the O'Neill game}
\end{flushleft}
According to this formula, for the O'Neill game,
Table \ref{tab:eigenval_vec_cyc} lists the eigencycle values
of the eigenvectors of the replicator dynamics.\\
\begin{flushleft}
\textbf{Interpretation of the definition}
\end{flushleft}
\begin{itemize}
\item \textbf{Invariance of the eigencycle:}
    An eigencycle is constructed by two components in a normalized eigenvector,
    and hence, its value is invariant.
    This is according to Eq.~(\ref{eq:the_ecyc}) and the following points:
    (1) the mode of each component is fixed;  and
    (2) the phase difference between two given components in an eigenvector is fixed;
    Because, they are modulated by the same eigenvalue,
    that is, the relative phase difference between these two components remains fixed.
\item \textbf{Number of eigencycle:}
    There are $N(N-1)/2$ independent eigencycles corresponding to
    a given $N$-component normalized eigenvector,
    as there are $N^2$ pairwise combinations of each component
    in an $N$-dimensional eigenvector.
    Considering that the $N$ self-combination of $(\eta_m,\eta_m)$ is trivial ($\sigma^{(mm)}$ = 0), and $(\eta_n,\eta_m)$ and $(\eta_m,\eta_n)$ are simply reversed ($\sigma^{(mn)}$ = $-\sigma^{(nm)}$), only $N(N\!-\!1)/2$ combinations remain.
\item \textbf{Eigencycle set:} The eigencycle set, as a vector denoted by $\Omega^{(mn)}_{\xi_k}$, is defined to represent a set of $N(N-1)/2$ eigencycle elements. The subscript is the normalized eigenvector $\xi$ indexed by $k$, which generates this eigencycle set. The superscript $(mn)$ is the index of the two-dimensional subspace, where the elements (eigencycles) of the set are located. $(mn)$ is defined as follows: \{$\{m,n\} \in \{1,2,...,n\} \cap (m < n)$\}. In this study, the assignment order is $m$ from 1 to $N$, and then $n$ from 2 to $N$.
\item \textbf{Subspace set:} The subspace set, denoted as $\Omega^{(mn)}$, has $N(N-1)/2$ eigencycle elements. The superscript $(mn)$ is the index of the two-dimensional subspace. Again, $(mn)$ is defined as follows: \{$\{m,n\} \in \{1,2,...,n\} \cap (m < n)$\}. In this study, the assignment order is $m$ from 1 to $N$, and then $n$ from 2 to $N$. Owing to symmetry, there exists a subset of $\Omega^{(mn)}$, wherein the performance of the elements is equal.
\item \textbf{Independence of eigencycle sets:} The eigencycle sets corresponding to eigenvectors are not all independent. Let us take the O'Neill game as an example. Among the eight eigencycle sets, only three (generated by $(\xi_{.8i},\xi_{.4i_1},\xi_{.4i_2})$) are independent. We explain this as follows:  Per the set of eigenvalues, (1) Two eigenvalues (0.2 and -0.2) are real numbers. As only the eigenvectors whose eigenvalues are complex numbers are related to cyclical motion, these two are trivial. (2) The remaining six eigenvectors make three pairs.
Each pair of eigenvectors with complex conjugate eigenvalues are conjugate,
and the generated eigencycle set pair has opposite values to the other pair. Hence, we can omit those related to eigenvalues, that is, $({-0.8i},{-0.4i},{-0.4i})$. Using the eigenvalues as the subscript symbols, the eigenvectors are recorded as $(\xi_{.8i},\xi_{.4i_1},\xi_{.4i_2})$.
%
In this system,
there two independent eigenvectors
(5-th and 7-th in Table \ref{tab:eigenval_vec_cyc}) belong to the same eigenvalue
${-0.4i}$.
We use the subscripts 1 and 2 to distinguish the eigencycle set from the two degenerate eigenvalues of $0.4i$.
\item \textbf{Geometric presentation:} A geometric presentation of an eigencycle set is shown in figure \ref{fig:Lissajous}.  An eigencycle only depends on the internal components $\eta_m$ and $\eta_n$ of the same normalized eigenvector. As both of the components' amplitude and phase difference is fixed and not arbitrary, so the geometric pattern set is fixed. In geometry, an elements of the eigencycle set is similar to a Lissajous cycle. An explanation of the similarity is shown in appendix section \ref{app:geo_eigc}.
\item \textbf{Relationship between eigencycle value and angular momentum:}
In mathematics, we can strictly prove that, the rate between the angular momentum at any time moment in any two subspace equals to  the rate between eigencycle values of the two subspace. See appendix section \ref{app:sigma_L} for details. Meanwhile, in a time series from a real experiment system, the angular momentum of the two observations in time series is measurable. So the eigencycle concept can be verified in real system.
\item \textbf{Why the O'Neill game?} Since O'Neill is a two-person four-strategy zero-sum game with eight dimensions, it is a high-dimensional problem. It is difficult for us to intuitively express the motion structure of such a high dimension. Using the eigencycle set, the eight-dimension game space can be decomposed into $28$ two-dimension sets. Importantly, \begin{itemize} \item In theory, this game's replicator dynamics equation can be explicitly solved mathematically. The game has high symmetry, so it is easily tractable. \item In experiment, the game matrix has high symmetry and there exist sufficient equivalent observations. The data should provide sufficient opportunity for testing. \item The O'Neill experiment was the first to reveal the reality and accuracy of the Nash Equilibrium in game theory. Revealing these characteristics  of game dynamics using our experiment data is a new challenge. \end{itemize}
\end{itemize}

\subsection{Explanation of the theoretical results}
Here, we briefly explain the theoretical results in Table \ref{tab:eigenval_vec_cyc}.
Meanwhile, define the fine structure and hypefine structure in the O'Meil game.
\begin{itemize}
\item \textbf{Precision and fine structure} Table \ref{tab:eigenval_vec_cyc} provides a theoretical prediction on cyclic behavior having higher precision than ever known. The eigencycle set $\sigma_{.8i}$ predicts four different values, (0.1964:0.0654:0.0218:0) = (9:3:1:0), that is, a cycle set with four different strengths are expected. This prediction leads to (1) the discovery of the fine structure (see section \ref{R2}); (2) the rising of the test accuracy one order (see section \ref{RE}), which are included in the main results of this study.
\item \textbf{Symmetry and hypefine structure} The contributions of the eigencycles from the different normalized eigenvectors $\xi$ are different, but can be classified as shown in the numerical results in Table \ref{tab:eigenval_vec_cyc}. The classification reflects the symmetry of the game matrix as well as the structure of the $\xi$ inner components. We employ this to identify the contribution of $\xi$ in the data (e.g., see $\sigma_\alpha$ and $\sigma_\beta$ in subsection \ref{R3}), which is one of  the main results of this study too.
\item \textbf{Operable ---} We use the experiment data to validate our eigencycle set decomposition approach. Our results may provide a spectrum analysis tool for game dynamics (see section \ref{R4}).
\end{itemize}

\begin{table}
\caption{The eigenvalues, eigenvectors, and their respective 28 eigencycles}
\begin{center}
\begin{tabular}{crrrrrrrrrr}
 \toprule 
 \tabincell{c}{Eigenvalue\\$\lambda_{i}$} ~~& $\lambda_{.8i}$ & $\lambda_{-.8i}$ & $\lambda_{.2}$ & $\lambda_{-.2}$ & $ \lambda_{.4i_1}$ & $\lambda_{-.4i_1}$ & $ \lambda_{.4i_2}$ & $\lambda_{-.4i_2}$ \\
 \hline
 \vspace{0.05cm}
	&$\frac{4}{5}\, \mathrm{i}$ & $-\frac{4}{5}\, \mathrm{i}$
	&$\frac{1}{5}$ & $-\frac{1}{5}$
	&$\frac{2}{5}\, \mathrm{i}_1$ & $-\frac{2}{5}\, \mathrm{i}_1$
	&$\frac{2}{5}\, \mathrm{i}_2$ & $-\frac{2}{5}\, \mathrm{i}_2$ \\
 \hline \vspace{0.1cm}
 \tabincell{c}{Eigenvector\\($\eta_i \in \xi$)} ~~& $\xi_{.8i}$ ~~~& $\xi_{-.8i}$ ~~~ & $\xi_{.2}$~~& $\xi_{-.2}$~~& $ \xi_{.4i_1}$~~ & $\xi_{-.4i_1}$~~~& $ \xi_{.4i_2}$~~ & $\xi_{-.4i_2}$\\
 \hline
$\eta_1$	&$-\frac{1}{4}\, \mathrm{i}$&$\frac{1}{4}\, \mathrm{i}$&$\frac{2}{5}$&0&	$	0 $&$ 0 $&$ 0 $&$ 0$\\
$\eta_2$	&$\frac{1}{12}\, \mathrm{i}$&$-\frac{1}{12}\, \mathrm{i}$&$\frac{1}{5}$&0&	$	\frac{1}{6} $&$ \frac{1}{6} $&$ \frac{1}{6} $&$ \frac{1}{6}$\\
$\eta_3$	&$\frac{1}{12}\, \mathrm{i}$&$-\frac{1}{12}\, \mathrm{i}$&$\frac{1}{5}$&0&	$	\frac{-1+\sqrt{3}\, \mathrm{i}}{12} $&$ \frac{-1-\sqrt{3}\, \mathrm{i}}{12} $&$ \frac{-1-\sqrt{3}\, \mathrm{i}}{12} $&$ \frac{-1+\sqrt{3}\, \mathrm{i}}{12}$\\
$\eta_4$	&$\frac{1}{12}\, \mathrm{i}$&$-\frac{1}{12}\, \mathrm{i}$&$\frac{1}{5}$&0&	$	 \frac{-1 -\sqrt{3}\, \mathrm{i}}{12} $&$ \frac{-1+\sqrt{3}\, \mathrm{i}}{12} $&$ \frac{ -1+\sqrt{3}\, \mathrm{i}}{12} $&$ \frac{-1-\sqrt{3}\, \mathrm{i}}{12}$\\
$\eta_5$	&$\frac{1}{4}$&$\frac{1}{4}$&0&$\frac{2}{5}$&	$	 0 $&$ 0 $&$ 0 $&$ 0$\\
$\eta_6$	&$-\frac{1}{12}$&$-\frac{1}{12}$&0&$\frac{1}{5}$&	$	 -\frac{\mathrm{i}}{6} $&$ \frac{\mathrm{i}}{6} $&$ -\frac{\mathrm{i}}{6} $&$ \frac{\mathrm{i}}{6}$\\
$\eta_7$	&$-\frac{1}{12}$&$-\frac{1}{12}$&0&$\frac{1}{5}$&	$	 \frac{\sqrt{3} + \mathrm{i}}{12} $&$ \frac{\sqrt{3}-\mathrm{i}}{12} $&$ \frac{-\sqrt{3}+\mathrm{i}}{12} $&$ \frac{-\sqrt{3}-\mathrm{i}}{12}$\\
$\eta_8$	&$-\frac{1}{12}$&$-\frac{1}{12}$&0&$\frac{1}{5}$&	$	 \frac{-\sqrt{3}+\mathrm{i}}{12} $&$ \frac{-\sqrt{3}-\mathrm{i}}{12} $&$ \frac{\sqrt{3}+\mathrm{i}}{12} $&$ \frac{\sqrt{3}-\mathrm{i}}{12}$\\
\hline
 \vspace{0.1cm}
\tabincell{c}{Eigencycle\\ ($\Omega^{(mn)}$)} & $\sigma_{.8i}$ ~~~& $\sigma_{-.8i}$ ~~~& $\sigma_{.2}$~~& $\sigma_{-.2}$~~& $ \sigma_{.4i_1}$~~ & $\sigma_{-.4i_1}$~~~& $ \sigma_{.4i_2}$~~ & $\sigma_{-.4i_2}$\\
 \hline
	 12 & 0 & 0 & 0 & 0 & 0 & 0 & 0 & 0 \\
	 13 & 0 & 0 & 0 & 0 & 0 & 0 & 0 & 0 \\
	 14 & 0 & 0 & 0 & 0 & 0 & 0 & 0 & 0 \\
	 15 & $-$0.1964 & 0.1964 & 0 & 0 & 0 & 0 & 0 & 0 \\
	 16 & 0.0654 & $-$0.0654 & 0 & 0 & 0 & 0 & 0 & 0 \\
	 17 & 0.0654 & $-$0.0654 & 0 & 0 & 0 & 0 & 0 & 0 \\
	 18 & 0.0654 & $-$0.0654 & 0 & 0 & 0 & 0 & 0 & 0 \\
	 23 & 0 & 0 & 0 & 0 & $-$0.0756 & 0.0756 & 0.0756 & $-$0.0756 \\
	 24 & 0 & 0 & 0 & 0 & 0.0756 & $-$0.0756 & $-$0.0756 & 0.0756 \\
	 25 & 0.0654 & $-$0.0654 & 0 & 0 & 0 & 0 & 0 & 0 \\
	 26 & $-$0.0218 & 0.0218 & 0 & 0 & 0.0873 & $-$0.0873 & 0.0873 & $-$0.0873 \\
	 27 & $-$0.0218 & 0.0218 & 0 & 0 & $-$0.0436 & 0.0436 & $-$0.0436 & 0.0436 \\
	 28 & $-$0.0218 & 0.0218 & 0 & 0 & $-$0.0436 & 0.0436 & $-$0.0436 & 0.0436 \\
	 34 & 0 & 0 & 0 & 0 & $-$0.0756 & 0.0756 & 0.0756 & $-$0.0756 \\
	 35 & 0.0654 & $-$0.0654 & 0 & 0 & 0 & 0 & 0 & 0 \\
	 36 & $-$0.0218 & 0.0218 & 0 & 0 & $-$0.0436 & 0.0436 & $-$0.0436 & 0.0436 \\
	 37 & $-$0.0218 & 0.0218 & 0 & 0 & 0.0873 & $-$0.0873 & 0.0873 & $-$0.0873 \\
	 38 & $-$0.0218 & 0.0218 & 0 & 0 & $-$0.0436 & 0.0436 & $-$0.0436 & 0.0436 \\
	 45 & 0.0654 & $-$0.0654 & 0 & 0 & 0 & 0 & 0 & 0 \\
	 46 & $-$0.0218 & 0.0218 & 0 & 0 & $-$0.0436 & 0.0436 & $-$0.0436 & 0.0436 \\
	 47 & $-$0.0218 & 0.0218 & 0 & 0 & $-$0.0436 & 0.0436 & $-$0.0436 & 0.0436 \\
	 48 & $-$0.0218 & 0.0218 & 0 & 0 & 0.0873 & $-$0.0873 & 0.0873 & $-$0.0873 \\
	 56 & 0 & 0 & 0 & 0 & 0 & 0 & 0 & 0 \\
	 57 & 0 & 0 & 0 & 0 & 0 & 0 & 0 & 0 \\
	 58 & 0 & 0 & 0 & 0 & 0 & 0 & 0 & 0 \\
	 67 & 0 & 0 & 0 & 0 & $-$0.0756 & 0.0756 & 0.0756 & $-$0.0756 \\
	 68 & 0 & 0 & 0 & 0 & 0.0756 & $-$0.0756 & $-$0.0756 & 0.0756 \\
	 78 & 0 & 0 & 0 & 0 & $-$0.0756 & 0.0756 & 0.0756 & $-$0.0756 \\
 \hline
\end{tabular}
\end{center}
 \vspace{-0.2cm}
\footnotesize{
Note: $\Omega^{(mn)}$ is the identity of the two-dimensional subspace of an eigencycle.
$m$ ($n$) indicates which among the eight dimensions
is selected as the $x$-axis ($y$-axis) in the eigencycle subspace.
}
\label{tab:eigenval_vec_cyc}
\end{table}

\section{Experimental observation on subspace cycles}
\subsection{Brief summary of the experiment data}
Table \ref{tab:datasource} summarizes the O?Neill game experiments. The three experiments span 26 years. In total, 358 subjects participated in the experiments, which involved multi-round repeated playing in the game matrix \ref{tab:gamemodel}.
The experiment data help verify the high-dimensional cycling in this study.

\begin{table}[!ht]
\caption{Experiment Data Source}
 \vspace{-0.2cm}
\footnotesize
\begin{center}
\begin{tabular}{|c|c|c|c|c|c|}
 \hline
\tabincell{c}{Experiment\\abbreviate}& \tabincell{c}{Data\\Source}&\tabincell{c}{Experiment\\Summary} &	\tabincell{c}{Published\\Time}&\tabincell{c}{total\\Rounds}&	\tabincell{c}{Matching\\Protocol}\\	\hline
$O$&O'Neill&\tabincell{c}{	The game was played by 50\\ students working in 25 pairs.}	&	1987&	2625&	\tabincell{c}{Fixed\\Paired}\\	\hline
$B$& Binmore et al&\tabincell{c}{		Each experiment session required 12 subjects \\13 experimental sessions in all.\\ Each real game was played 150 times.}&	2001 &	1950&	\tabincell{c}{Randomly\\Matching}\\	\hline
$IT$&	 Okano&\tabincell{c}{20 individuals adopted the player A role\\ against 20 teams adopting the player B role.\\ 20 experimental sessions in all.\\ Each real game was played 150 times.}&	2013&	3000&	\tabincell{c}{Fixed\\Paired}\\	\hline
$TI$&	 Okano&\tabincell{c}{20 teams adopted the player A role \\against 20 individuals adopting \\the player B role, 20 experimental sessions \\in all. Each real game was played 150 times.}&	2013&	3000&	\tabincell{c}{Fixed\\Paired}\\	\hline
$II$&	 Okano&\tabincell{c}{Individuals against individuals, \\18 experimental sessions in all. \\Each real game was played 132 times.}&	2013&	2376&	\tabincell{c}{Fixed\\Paired}\\	\hline
$TT$&	 Okano&\tabincell{c}{Teams against teams, \\18 experimental sessions in all. \\Each real game was played 132 times.}&	2013&	2376&	\tabincell{c}{Fixed\\Paired} \\	\hline
\end{tabular}
\end{center}
\label{tab:datasource}
\end{table}

\subsection{Angular momentum as the measurement}

\begin{flushleft}
\textbf{The measurement for the cycle in the subspace: }
\end{flushleft}

According to the theoretical eigencycle set decomposition approach,
we can measure the cyclic angular momentum
~\footnote{This measurement is the signed area of
the triangle $\Delta_{[O, x(t), x(t+1)]}$ in the ($m,n$) two-dimensional subspace.
For each transition from $x(t)$ to $x(t+1)$, referring to $O$,
the angular momentum is twice the signed area of the triangle.
We suggest using the angular momentum
because it contains
the mass $m$ as a parameter,
which may be compatible with the population size $N$
as the variable in further investigations of game dynamics. }
in each of the two-dimensional subspace,
indicated by the eigencycle $\Omega^{(mn)}$, separately.
The angular momentum $L^{(mn)}_E$\cite{2012Inertia} can be expressed by the following formula: \\
\begin{equation}\label{eq:exp_am} L^{mn}_E=\frac{1}{N-1}\sum_{t=1}^{N-1}\left(x(t)-O\right) \times \left(x(t\!+\!1)-x(t)\right)
\end{equation}

\begin{itemize}
\item $L^{(mn)}_E$ represents the average value of
the accumulated angular momentum over time;
the subscript $mn$ indexes the two-dimensional $(x_m,x_n)$ subspace;
\item $N$ is the length of the experimental time series, that is,
the total number of repetitions of the repeated game experiments;
\item $O$ is the projection of the Nash equilibrium at the subspace $\Omega^{(m,n)}$;
\item $x(t)$ is a two-dimensional vector at time $t$,
which can be expressed as $(x_m(t),x_n(t))$, $x(t+1)$ is a two-dimensional vector at time $t+1$, and
\item $\times$ represents the cross product between two two-dimensional vectors.
\end{itemize}

\begin{flushleft}
\textbf{Interpretation of the measurement: }
\end{flushleft}

\begin{itemize}
\item The sign of the value of $L^{(mn)}_E$ indicates the direction of the cyclical movement (counterclockwise or clockwise): $L^{(mn)}_E >0$ refers to counterclockwise movement and $L^{(mn)}_E < 0$ refers to clockwise movement. $L^{(mn)}_E =0 $ means no cycle was observed. If a system is completely random, as predicted by the mixed strategy Nash equilibrium, the long-term average of its angular momentum is 0.
\item The modulus of the value, $||L^{(mn)}_E||$, indicates the strength of determinate motion. For example, in a one-to-one fixed pairing, there are four states in a two-dimensional subspace, and these four states form a unit square with an area of 1. The accumulated angular momentum of one full cycle is 2, or $||L^{(mn)}_E||=2$.
\item If, in the time series of 1,000 repeated rounds experiment, the observed accumulated $L^{(mn)}_E = 40$, then the total amount of the determinate motion is 20 cycles. This also means there are 80 determinate motions beyond complete stochastic motion, that is, the probability of directional movement is 8\%.
\end{itemize}

\subsection{Experimental results}

We calculate the average angular momentum for the experiments shown in Table \ref{tab:datasource}, and Table \ref{tab:exp_am} shows the results.

\begin{table}
\caption{Experimental angular momentum}
 \vspace{-0.2cm}
\begin{center}
\begin{tabular}{ c r r r r r r }
 \toprule
 \vspace{0.1cm}
$\Omega^{(mn)}$&	$L^{(mn)}_O$~~&	$L^{(mn)}_B$~~&	$L^{(mn)}_{IT}$~~&	$L^{(mn)}_{TI}$~~&	$L^{(mn)}_{II}$~~&	$L^{(mn)}_{TT}~~$\\\hline
12	&	0.003	&	0.002	&	$-0.008$	&	$-0.001$	&	0.002	&	0.008	\\	\hline
13	&	$-0.011$	&	$-0.002$	&	0.003	&	0.000	&	$-0.009$	&	$-0.002$	\\	\hline
14	&	0.008	&	0.000	&	0.004	&	0.002	&	0.007	&	$-0.006$	\\	\hline
15	&	$-0.038$	&	$-0.015$	&	$-0.053$	&	$-0.033$	&	$-0.039$	&	$-0.049$	\\	\hline
16	&	0.009	&	0.010	&	0.018	&	0.019	&	0.008	&	0.020	\\	\hline
17	&	0.021	&	0.002	&	0.020	&	0.006	&	0.015	&	0.014	\\	\hline
18	&	0.007	&	0.003	&	0.015	&	0.008	&	0.016	&	0.014	\\	\hline
23	&	0.011	&	0.003	&	$-0.009$	&	0.001	&	0.008	&	0.007	\\	\hline
24	&	$-0.008$	&	$-0.001$	&	0.001	&	$-0.002$	&	$-0.006$	&	0.001	\\	\hline
25	&	0.010	&	0.003	&	0.021	&	0.005	&	0.015	&	0.018	\\	\hline
26	&	$-0.001$	&	0.000	&	0.004	&	$-0.011$	&	0.002	&	$-0.009$	\\	\hline
27	&	$-0.001$	&	$-0.001$	&	$-0.010$	&	0.004	&	$-0.012$	&$	-0.014$	\\	\hline
28	&	$-0.007	$&	$-0.002$	&	$-0.014	$&	0.002	&	$-0.005$	&	0.005	\\	\hline
34	&	0.000	&	0.001	&	$-0.006$	&	0.000	&	$-0.001$	&	0.005	\\	\hline
35	&	0.011	&	0.008	&	0.015	&	0.011	&	0.014	&	0.013	\\	\hline
36	&	$-0.009$	&	$-0.006$	&$-0.003$	&	$-0.002$	&	$-0.006$	&	$-0.005$	\\	\hline
37	&	$-0.003$	&	0.001	&	$-0.004$	&	$-0.004$	&	0.002	&	0.002	\\	\hline
38	&	0.002	&	$-0.002$	&	$-0.008$	&	$-0.006$	&	$-0.011$	&	$-0.010$	\\	\hline
45	&	0.018	&	0.004	&	0.017	&	0.017	&	0.010	&	0.019	\\	\hline
46	&	0.000	&	$-0.004$	&	$-0.019$	&	$-0.006$	&	$-0.003$	&	$-0.007$	\\	\hline
47	&	$-0.016$	&	$-0.001$	&	$-0.006$	&	$-0.007$	&	$-0.006$	&	$-0.002$	\\	\hline
48	&	$-0.002$	&	0.002	&	0.007	&	$-0.004$	&	$-0.001$	&	$-0.009$	\\	\hline
56	&	$-0.006$	&	0.001	&	$-0.001$	&	0.001	&	$-0.001$	&	0.002	\\	\hline
57	&	0.009	&	0.000	&	0.004	&	0.005	&	$-0.005$	&	$-0.009$	\\	\hline
58	&	$-0.003$	&	$-0.001$	&	$-0.002$	&	$-0.006$	&	0.006	&	0.007	\\	\hline
67	&	$-0.003$	&	0.001	&	$-0.003$	&	0.000	&	0.005	&	0.007	\\	\hline
68	&	$-0.003$	&	0.000	&	0.002	&	0.001	&	$-0.006$	&	$-0.005$	\\	\hline
78	&	0.006	&	0.000	&	0.000	&	0.005	&	0.000	&	$-0.002$	\\	\hline
\end{tabular}
\end{center}
 \vspace{-0.2cm}
\footnotesize{
$\Omega^{(mn)}$ is the $28$ subspaces referring to Table
\ref{tab:eigenval_vec_cyc}; the subscript $L$ refers to the O'Neill and Binmore as well as the AIBT, ATBI, AIBI, and
ATBT experiments in Table \ref{tab:datasource}.
We use the software MATLAB, version R2018a.}\\

\label{tab:exp_am}
\end{table}

\begin{table}
\caption{Consistency of the six experiments}
 \vspace{-0.2cm}
 \begin{center}
\begin{tabular}{crrrrrr}
 \toprule
 	&~~~~$L_O$~~~&~~~~$L_B$~~~&~~~$L_{IT}$~~~	&~~~$L_{TI}$~~~&~~~$L_{II}$~~~&~~~$L_{TT}$~~~\\ \hline
 $L_O$ & 1.000 & & & & & \\ &0.000& & & & & \\\hline
$L_B$ & 0.706 & 1.000 & & & & \\
 & 0.000 &0.000 & & & & \\\hline
$L_{IT}$ & 0.522 & 0.639 & 1.000 & & & \\
 & 0.004 & 0.000 & 0.000 & & & \\\hline
$L_{TI}$ & 0.660 & 0.658 & 0.586 & 1.000 & & \\
 & 0.000 & 0.000 & 0.001 &0.000 & & \\\hline
$L_{II}$ & 0.728 & 0.825 & 0.608 & 0.5306 & 1.000 & \\
 & 0.000 & 0.000 & 0.001 & 0.0037 &0.000 & \\\hline
$L_{TT}$ & 0.474 & 0.756 & 0.470 & 0.5932 & 0.770 & 1.000\\
 & 0.011 & 0.000 & 0.012 & 0.0009 & 0.000 &0.000 \\\hline
\end{tabular}
\end{center}
 \vspace{-0.2cm}
\footnotesize{
Measure: Spearman's rank correlation.
The number of observations is 28.
The variables (observations) are the experimental angular momentum
shown in Table \ref{tab:exp_am}.
The first row reports the correlation coefficients, and
the second row reports the significance level $p$.
We use the software Stata, version:SE 15.1.}
\label{tab:spearman}
\end{table}

\subsection{Explanation of the experimental results:\label{RE}}
The measurement results illustrate the consistency and precision
 of the six experiments, thus validating the inner eigenvector decomposition approach. The cycles in the high-dimension space are real and testable as well, as shown by the following two points:
\begin{itemize}
\item \textbf{Consistency~------}
    In the six experiments, the observed cycle values $L$ in the two-dimensional subspaces show consistency (Spearman rank's test, max($p$)=0.012, mean of the $\rho$ values is 0.6351, Std.~\!Ev = 0.1087, Std.~\!Err = 0.0118).
\item \textbf{Precision~------}
Except for the main cycle found in $\Omega^{(15)}$, we can now test the cycles in the subspace ($\Omega^{(26,27,28,36,37,38,46,47,48)}$). This means that the accuracy increases up to one order of magnitude. We explain this with two numerical examples: \begin{itemize}
\item $L^{(15)}_O$=$-0.038$, where the superscript $O$ indicates the O'Neill game, superscript $(15)$ indicates that the subspace measured is $\Omega^{(15)}$, and the negative sign means the cycle is clockwise. $||L||$ = 0.038, that is, there are ($\frac{105\times 0.038}{2}\simeq$ 2) two cycles obtained in the O'Neill (1987) 105-rounds game, and, in the 25 groups, 2,625 rounds of observations, yielding  50 cycles.
\item $L^{(37)}_O$=$-0.003$, that is, the average angular momentum in subspace 37 is 0.003. Each complete cyclical motion in the four-states subspace of the O'Neill game contributes two units of angular momentum; then, the 105 rounds have 0.2 cycles, and the 2,625 rounds have 4.4 cycles.
\end{itemize}
\end{itemize}

\section{Eigencycle set verified with experiments}
We have shown the theoretical
eigencycle values $\sigma^{(mn)}$ in Table. \ref{tab:eigenval_vec_cyc} and the experimental average angular momentum $L_{mn}$ in Table \ref{eq:exp_am}.

There is following proposition ---
For various subspace $(m,n)$,
the average accumulated angular momentum
$L_{mn}$ divided by its $\sigma^{(mn)}$ is invariant
(Details of the mathematical proof is provided in appendix \ref{app:sigma_L}).
That is to say, by take $\sigma^{(mn)}$ as predictor (independent variable) and take $L_{mn}$
as the observation (dependent variables),  a linear relation with truncation term being 0 can be expected.
So, we can evaluate the validity of the eigencycle with the experimental $L$ (average angular momentum).

\begin{figure}[!ht]
\centering
\includegraphics[scale=0.55]{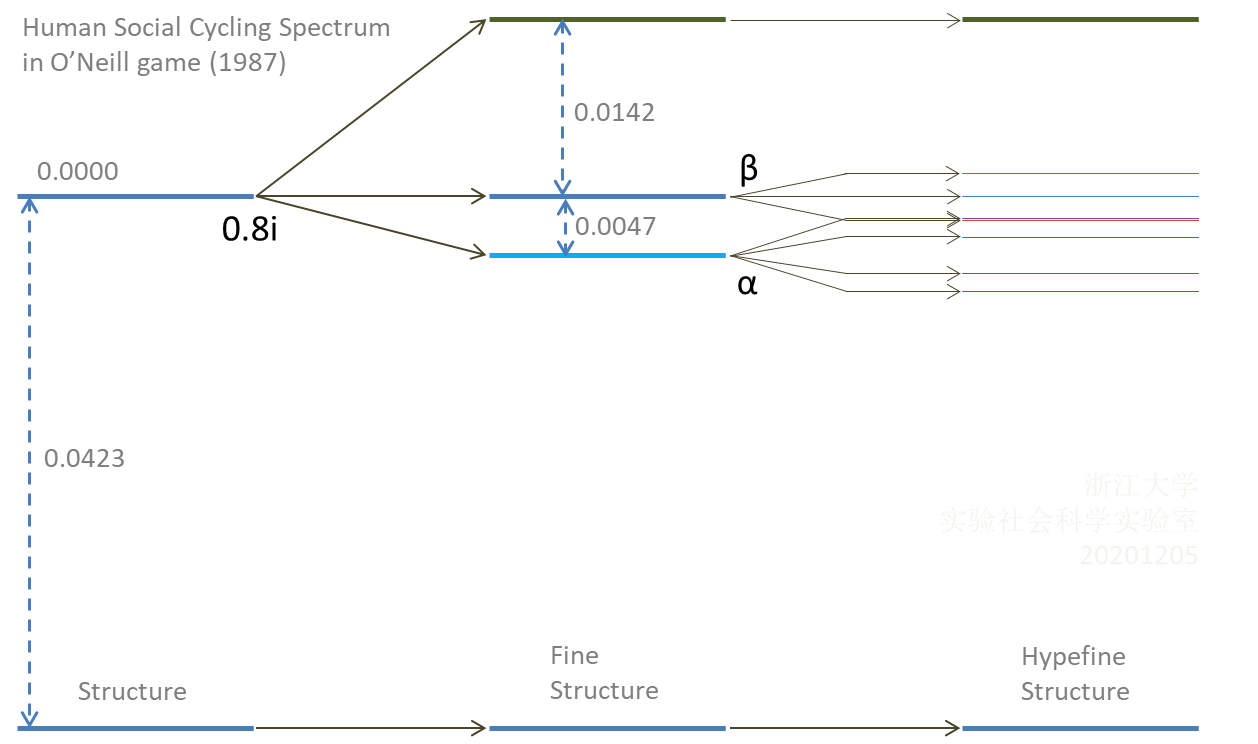}
\caption{Fine and hyperfine structures
of game dynamics in the O'Neill (1987) game experiments.
The labels \{0.0423, 0.0142, and 0.047\} are the observed average angular momentum of the experiments, which linearly match the eigencycle values of $\sigma_{.8i}$.
This figure illustrates how the theoretical eigencycle set
 clearly shows the existence of the
the fine and hyperfine structures, thus enhancing the literature
\cite{Binmore2001Minimax,2012arXiv1203.2591W,dan2014,dan2020}.}
\label{fig:fine_structure}
\end{figure}

The three subsections can be abstracted as follows:
\begin{enumerate}
\item Report the discovery of the fine structure, because of the contribution of $\sigma_{.8i}$ which is
labeled as $0.8i$ in Figure \ref{fig:fine_structure};
\item Report the finding of the hyperfine structure, because of the contribution of $\sigma_{.4i_1}$ and $\sigma_{.4i_2}$ which is
labeled as $\alpha$ and $\beta$ in Figure \ref{fig:fine_structure}, respectively;
\item Report the multi-OLE results (see Table \ref{tab:multi_reg}) illustrate that
the eigencycle set is an ideal
basis in this study.
The results validate the eigenvector set as a spectrum analysis tool on dynamics system.
\end{enumerate}

\subsection{Fine structure \label{R2}}
\begin{flushleft}
\textbf{Result 1's description: }
\end{flushleft}

\begin{figure}[!ht]
\centering
\includegraphics [width=1.87in]{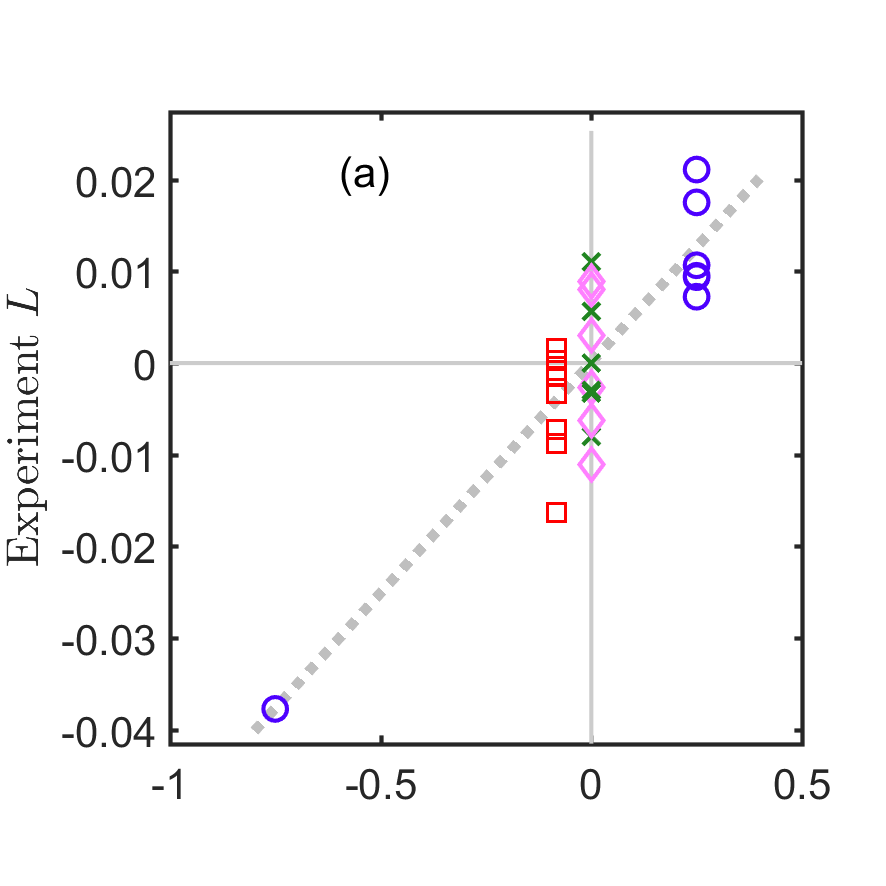}
\includegraphics [width=1.8in]{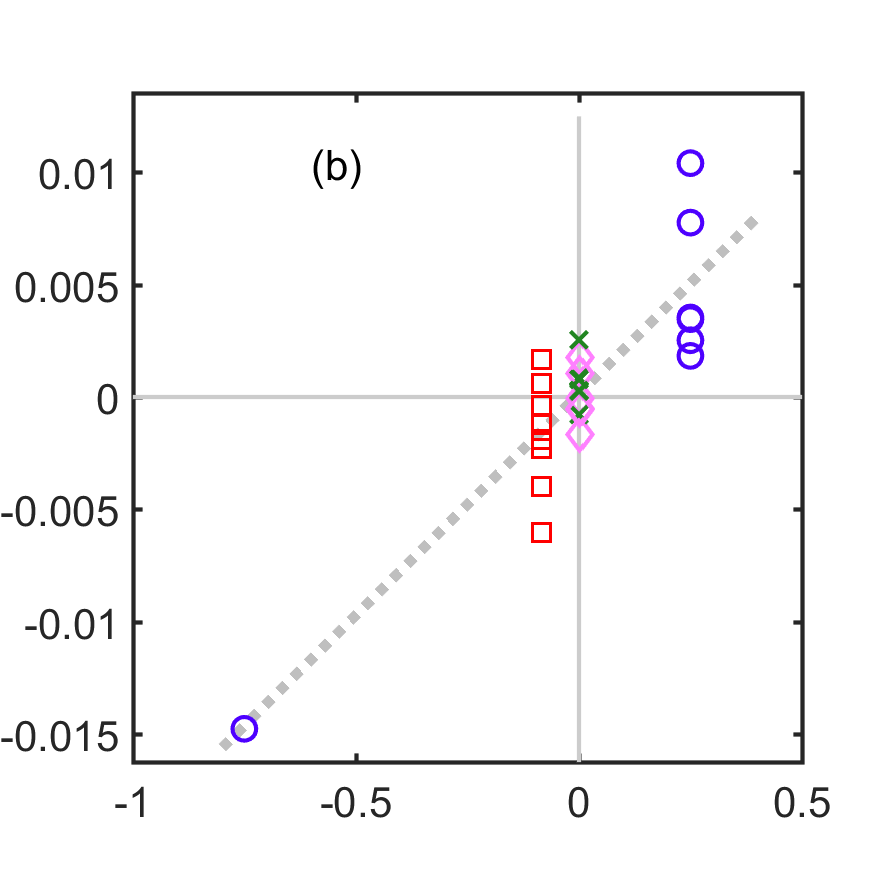}
\includegraphics [width=1.8in]{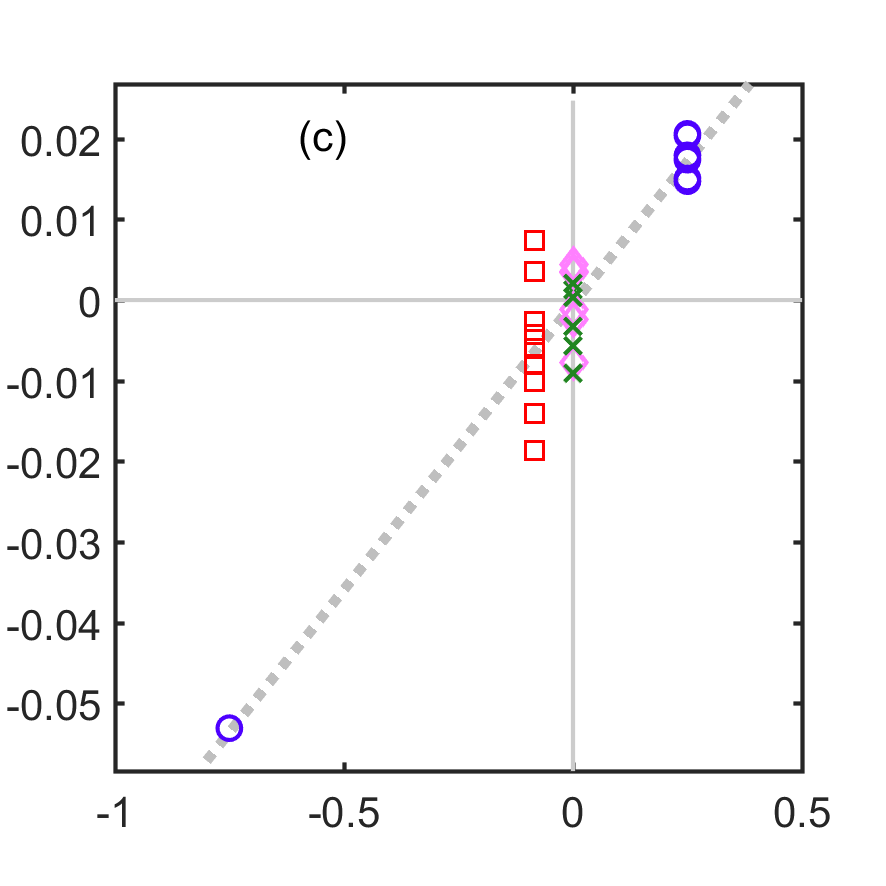}\\
\includegraphics [width=1.85in]{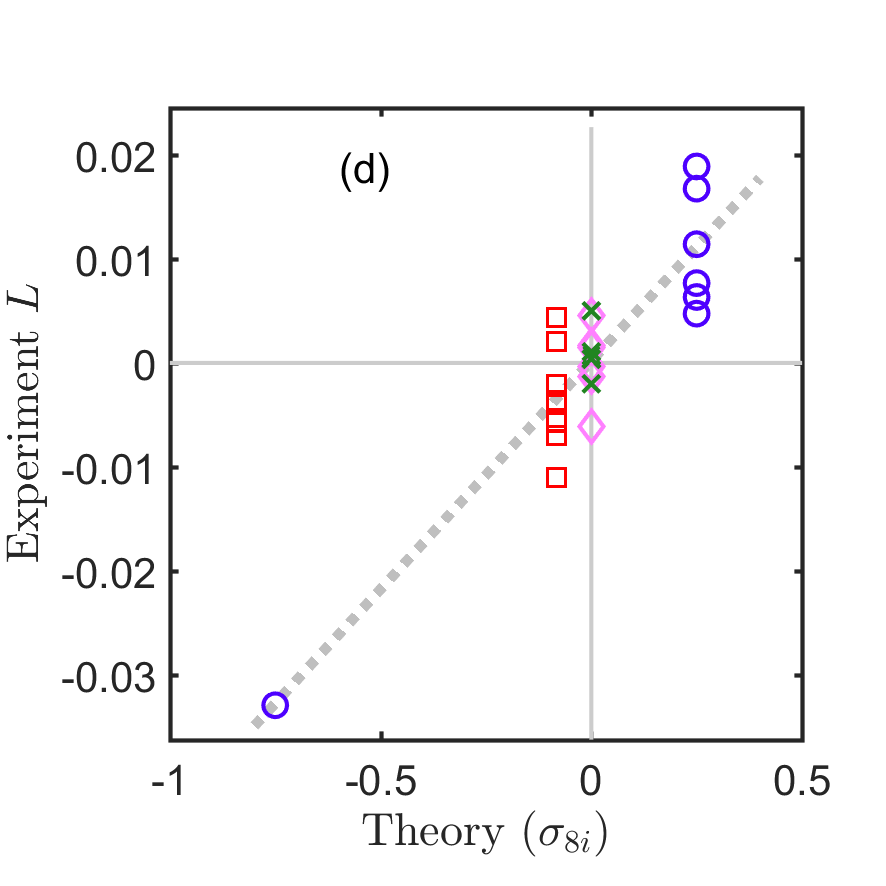}
\includegraphics [width=1.8in]{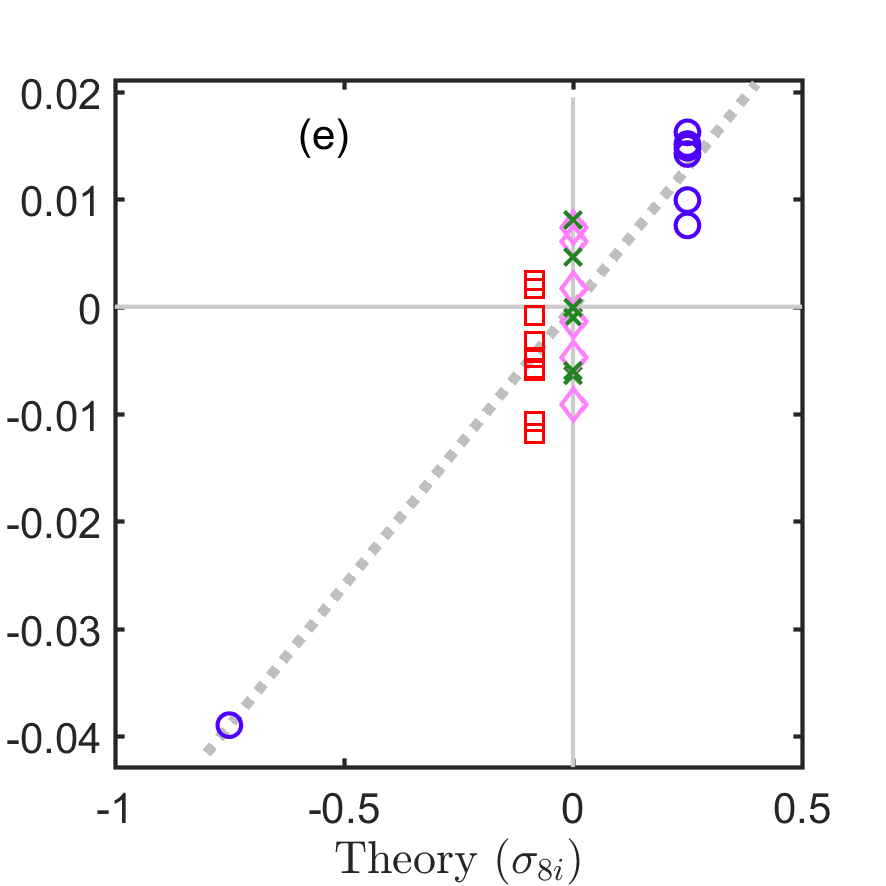}
\includegraphics [width=1.8in]{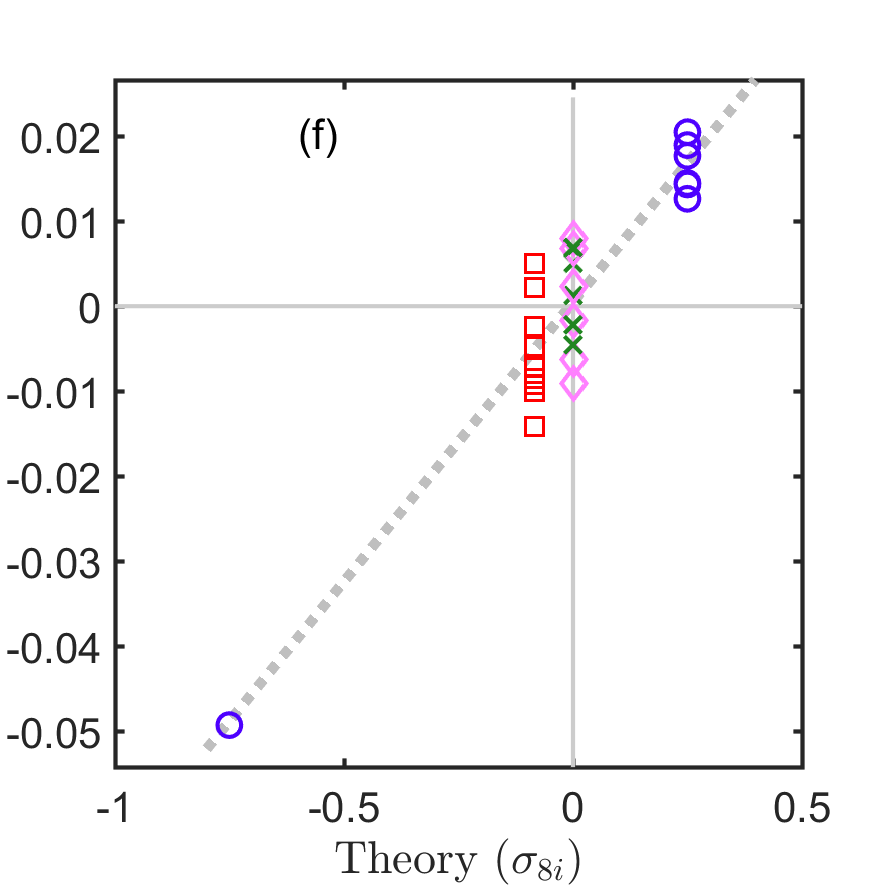}
\caption{Linear regression to test the validity of the eigencycle set concept with $\sigma_{.8i}$ through the experiment. The 9:3:1:0 predicted by $\sigma_{.8i}$ is observed, which leads to the discovery of the fine structure of game dynamics. The experiment values for (a)--(f) are from the treatments ordered as in Table \ref{tab:datasource}.}
\label{fig:8iregression}
\end{figure}

The six experiments' cycles
from the 28 subspaces are employed
to test the prediction of the three eigencycle sets
($\sigma_{.8i}$, $\sigma_{.4i_1} $, and $\sigma_{.4i_2}$)
by OLE, respectively.
%
The results from the six experiments are consistent:
\begin{itemize}
\item Among the 28 samples from the theoretical $\sigma_{.8i}$ (eigencycle set) and
experimental $L$, there exists a significant linear dependence.
The results show that $\sigma_{.8i}$ can independently be the
principle component for the game dynamics.
\item Importantly, we clearly observe
the four cycles values, rated as (9:3:1:0)
and predicted by $\sigma_{.8i}$ (see the red squares in Figure \ref{fig:8iregression}), indicating the discovery of the \textbf{fine structure}
in the O'Neill game.
\item 
Neither $\sigma_{.4i_1}$ nor $\sigma_{.4i_2}$ exists in significant linear dependence with experimental $L$.
for the game dynamics. However, they are not invalid, as we later show that they lead to the \textbf{hyperfine structure}.
\end{itemize}
%
%
\begin{flushleft}
\textbf{The supporting data of result 1: }
\end{flushleft}
The report of 'the discovery of the fine structure' is based on the
consequence of the $\sigma_{.8i}$
by splitting the $\Omega^{{15}}$ cycle in 5 $\sigma$ significant.
This claim is supported by the data as following:

Figure \ref{fig:8iregression} illustrates the ordinary least squares of the theoretical $\sigma_{.8i}$ and the six experimental $L$. Table \ref{tab:reg8i4i14i2} provides the results of the ordinary least squares between the six experiments and $\{\sigma_{.8i}, \sigma_{.4i_1}, \sigma_{.4i_2}\}$.
\begin{itemize}
\item On the claim of the discovery of the fine structure --- The six subspaces of the six experiments, when observed together, yield 36 samples, and the result is strongly significant ($ttest$, $p$=2.835 $\times 10^{-15}$, $N$=36).
More strictly,  considering the concentration of the 8 dimension,
omitting the 4- and 8-th dimension related subspaces, the result is strongly significant remained ($ttest$, $p$=1.3802 $\times 10^{-10}$, $N$=24).  This stratifies the statistical significance of the five standard deviations (5 $\sigma$, a threshold of $p \leq 2.87 \times 10^{-7}$) above background expectations, at 0. So, we claim the discovery of fine structure.
\item Regarding the fine structure,
the nine subspaces of the six experiments are observed together,
giving us 54 samples. In this case, the result is
strongly significant ($ttest$, $p$=7.689 $\times 10^{-7}$, $N$=54).
We are not suprise by this relatively low significant,
because the theoretical value is closer to background expectations, at 0.
\end{itemize}

\begin{table}
\caption{Ordinary least squares between the six experimental $L$ and $\{\sigma_{.8i}, \sigma_{.4i_1}, \sigma_{.4i_2}\}$} \small
\begin{center}
\begin{tabular}{crccccccc}
 \toprule
& ~~~~~~&$L^{(mn)}_O$&$L^{(mn)}_B$& $L^{(mn)}_{IT}$& $L^{(mn)}_{TI}$& $L^{(mn)}_{II}$& $L^{(mn)}_{TT}$ \\ \hline
$\sigma_{.8i}$&coef.~ $t$&	8.26&	9.31&	12.78&	10.72&	10.64&	12.39 \\	
&$p$&	(0.00)&	(0.00)&	(0.00)&	(0.00)&(0.00)&	(0.00) \\	
&cons.~ $t$&0.08&	0.35&	-0.49&	0.25&	-0.03&	0.47 \\	
&$p$&(0.939)&	(0.726)&	(0.630)&	(0.801)&	(0.978)&	(0.640) \\	
&$R^2$&	0.724&	0.7692&	0.8627&	0.8155&	0.8131&	0.8552 \\	\hline
$\sigma_{0.4i_1}$&coef.~$t$&	$-0.31$&	0.47&	1.24&	$-0.63$&	0.11&	$-0.36$ \\	
&$p$&	(0.759)&	(0.645)&	(0.227)&	(0.535)&	(0.917)&	(0.719) \\	
&cons.~t&	0.01&	0.22&	$-0.05$&	0.04&	0&	0.14 \\	
&$p$&	(0.995)&	(0.827)&	(0.960)&	(0.968)&	(0.999)&	(0.89) \\	
&$R^2$&	0.0037&	0.0083&	0.0557&	0.0149&	0.0004&	0.0051 \\	\hline
$\sigma_{.4i_2}$&coef.~$t$&	0.92&	1.11&	0.43&	$-0.2$&	1.39&	0.42 \\
& ~$p$&	(0.367)&	(0.276)&	(0.672)&	(0.843)&	(0.176)&	(0.675) \\	
&cons.~$t$&	$-0.06$&	0.05&	$-0.23$&	0.13&	$-0.17$&	0.13 \\	
&p&	(0.952)&(0.960)&(0.822)&	(0.898)&	(0.870)&	(0.896) \\	
&$R^2$&	0.0314&	0.0454&	0.007&	0.0015&	0.0692&	0.0069 \\	\hline
$\sigma_{.8i}=-0.0833$&mean&$-0.0042$&	$-$0.0016&	$-$0.0059&	$-0.0037$&	$-$0.0043&	$-$0.0055 \\	
&$\pm SE$&(0.0019)&	(0.0008)&(0.0027)&	(0.0016)&	(0.0017)&	(0.0021) \\	\hline
\end{tabular}
\end{center}
\vspace{-0.2cm}
\footnotesize{Software:Stata, version:SE 15.1}
\label{tab:reg8i4i14i2}
\end{table}

\begin{flushleft}
\textbf{Interpretation of result 1:}
\end{flushleft}
\begin{itemize}
\item \textbf{$\sigma_{.8i}$ as principal component~---~}
The $p$ value of the regression coefficient of $\sigma_{.8i}$ is close to $0.000$, and the constant term cannot reject the $0$ hypothesis. Thus, $\sigma_{.8i}$ is explanatory with respect to the main experimental results. We can regard $\sigma_{.8i}$ as the principal component in this system. $\sigma_{.8i}^{(15)}$ has the largest value, which is consistent with Binmore et al.?s suggestion \cite{Binmore2001Minimax} and confirmed by Wang and Xu \cite{2012arXiv1203.2591W} after converting the $4\times 4$ game into a $2 \times 2$ payoff matrix. To our knowledge, the observed cycles in $\sigma_{.8i}^{(15)}$ are cutting-edge contributions on cycle measurement in the literature.
\item \textbf{Discover the fine structure~---~} 
Beyond the $\Omega^{(15)}$ subspaces, $\sigma_{.8i}$ also predicts the cycle 
in the six subspaces, $\Omega^{(16,17,18,25,35,45)}$, and
the nine two-dimensional subspaces, $\Omega^{(26,27,28,36,37,38,46,47,48)}$.  These cycles are one-third or one-ninth in strength compared with $\Omega^{(15)}$'s cycle, which cannot be tested by earlier methods. However, the eigencycle set approach shows their significant existence. We call these the 'fine structure'.
\item \textbf{Interpretation of $\sigma_{.4i}$ results~---~} %
The no-zero eigencycles in the $\sigma_{.4i_1}$ and $\sigma_{.4i_2}$ set are all independent of the first or fifth dimension. Hence, the experimental observed cycles in $\Omega^{(15,16,17,18,25,35,45)}$, which are obvious motions by $\sigma_{.8i}$, cannot be captured. As a result, neither $\sigma_{.4i_1}$ nor $\sigma_{.4i_2}$ can globally capture the entire 28-subspace motion reasonably.
\end{itemize}

\subsection{Hyperfine structure \label{R3}}
\begin{flushleft}
\textbf{Result 2's description}
\end{flushleft}
We investigate $\sigma_{.4i_1}$ and $\sigma_{.4i_2}$
using the data.
As per the main results, we find the hyperfine structure.
In $\sigma_{.4i_1}$ and $\sigma_{.4i_2}$, the corresponding eigenvectors ($\xi_{.4i_1}$ and $\xi_{.4i_2}$) and components
$\eta_1$ and $\eta_5$ are all 0. Hence,
$\sigma_{.4i_1}$ and $\sigma_{.4i_2}$
have an effect only in the (2-3-4, 6-7-8) six-dimension subspace,
which includes 15 two-dimensional subspaces
where the eigencycles can exist.
We use the two degenerate eigenvalues' eigencycle set
($\sigma_{.4i_1}$, $\sigma_{.4i_2}$)
to build a pair of orthogonal bases
as follows:
\begin{equation}\label{eq:alphabeta}
\left\{ \begin{array}{lr} \sigma_\alpha= (\sigma_{.4i_2}+\sigma_{.4i_1})/2, \\ \sigma_\beta = (\sigma_{.4i_2}-\sigma_{.4i_1})/2 \end{array}
\right.
\end{equation}
The mathematical explanations for the two bases are as follows:
\begin{itemize}
\item $\sigma_\alpha$ only relates to the (2-3-4, 6-7-8)-dimension outer-cross nine subspaces, $\Omega^{(26,27,28,36,37,38,46,47,48)}$. That is,
$\sigma_\alpha$ has only nine components, but fully covers the nine two-dimensional subspace. Notably, as a strictly mathematical result, there exist only two values, and they are in opposition in $\sigma_\alpha$.
\item $\sigma_\beta$ only relates to the (2-3-4, 6-7-8)-dimension inner-cross six subspaces, $\Omega^{(23,24,34,67,68,78)}$. That is,
$\sigma_\alpha$ has only six components, but fully covers the six two-dimensional subspaces.
Again, there exists only one value, and it is in $\sigma_\beta$.
\end{itemize}
The results of the experimental test are reported as follows:
\begin{itemize}
\item There exists significant linear dependence
between the experimental $L$ and $\sigma_\alpha$. The constant term is also a consequence of
$\sigma_{.8i}$. Hence, we have discovered the hyperfine structure.
\item
There exists significant linear dependence
between the experimental $L$ and $\sigma_\beta$. However, the constant term is zero and cannot be rejected.
\end{itemize}

\begin{figure}[!ht]
\centering
\includegraphics [width=1.87in]{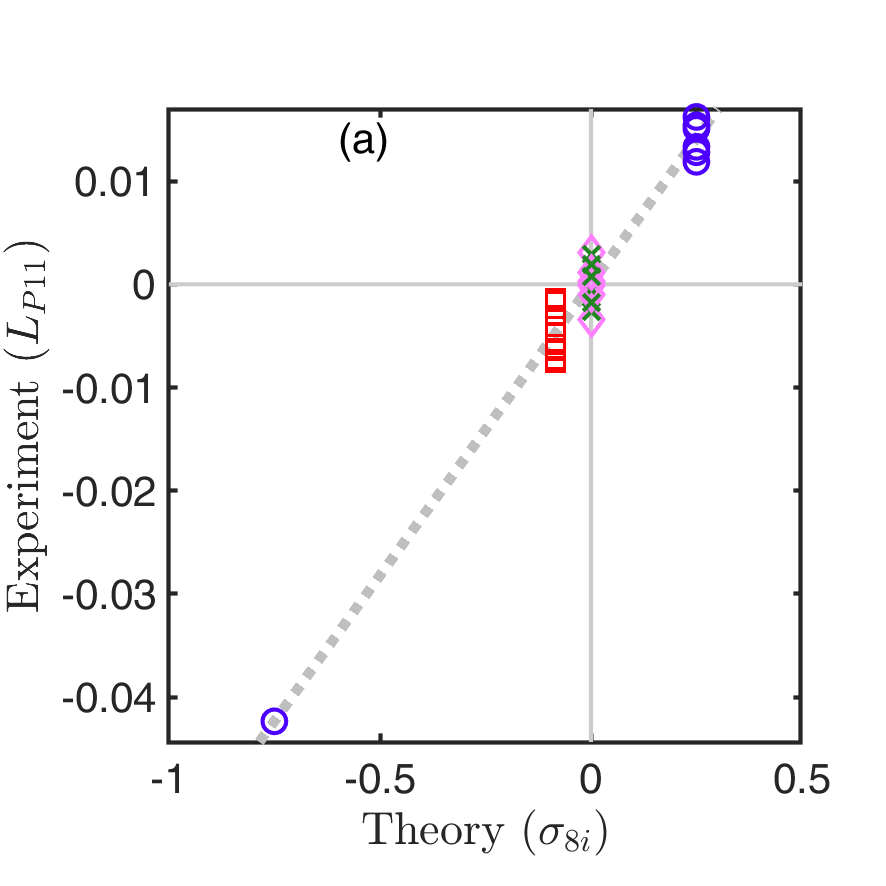}
\includegraphics [width=1.8in]{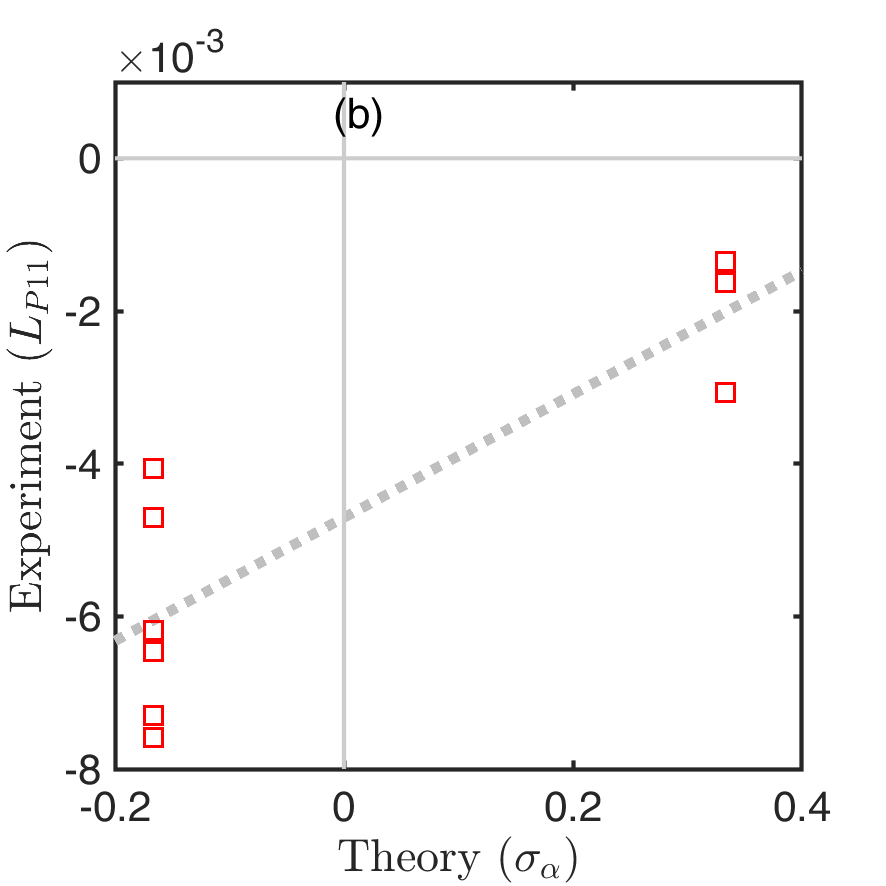}
\includegraphics [width=1.8in]{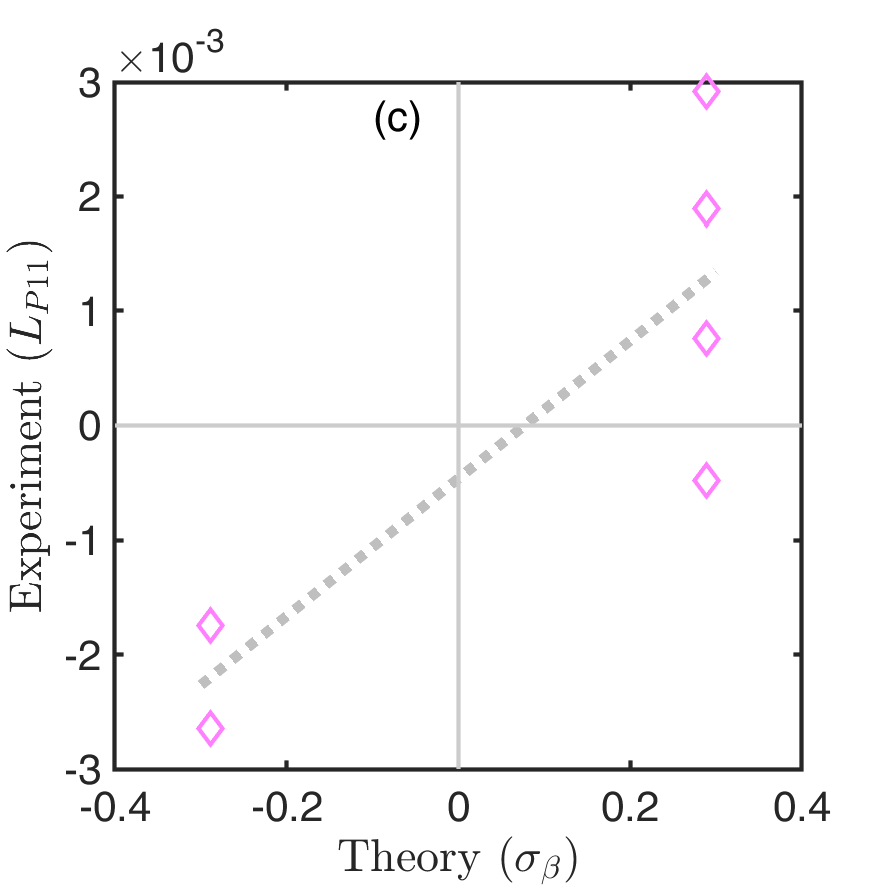}\\
\caption{Fine structure and hyperfine structure of the cycles}
\label{fig:lp11_reg}
\end{figure}

\begin{flushleft}
\textbf{Supporting data of result 2}
\end{flushleft}
The report of 'find evidence of hyperfine structure' is based on the
statistical significant consequence of $\sigma_\alpha$ and  $\sigma_\beta$ on
the fine structure due to $\sigma_{.8i}$.
The supporting data are as following:
\begin{itemize}
     \item Table \ref{tab:P11_alpha_beta} and Figure \ref{fig:lp11_reg} show the supporting data. As the five treatments (labeled ($O, IT, TI, II, and~TT$) in Table \ref{tab:datasource})
    have the same social state pattern (each two-dimensional subspace has four states), we thus pool the data under the label $P11$-treatment.
    \item By regressing $L_{P11}$ and $\sigma_{.8i}$ as well as $\sigma_{\alpha}$ and $\sigma_{\beta}$, we can obtain the result of the first, second, and third rows in Table \ref{tab:P11_alpha_beta}, respectively, as shown in Figure \ref{fig:lp11_reg}.
        \begin{itemize}
        \item $L_{P11}$ and $\sigma_{.8i}$ have significant positive correlation (linear regression, $t$=27.36, p=0.000, N=28);
        See Figure \ref{fig:lp11_reg}(b).
        \item $L_{P11}$ and $\sigma_{\alpha}$ have significant positive correlation (linear regression, $t$=4.43, $p$=0.003, N=9), wherein the constant item is the influence of $0.8i$ on the nine points ($0.057\times(-0.083)=-0.0047\in\{-0.006,-0.004\}$);
        See Figure \ref{fig:lp11_reg}(b).
        \item $L_{P11}$ and $\sigma_{\beta}$ have significant positive correlation (linear regression, $t$=3.31, $p$=0.030, N=6);
        See Figure \ref{fig:lp11_reg}(c).\\
        \end{itemize}
    \item If sample by the experiments and the subspace simultaneously, we have larger sample size, and we get the same result.
    \begin{itemize}
        \item If the $\sigma_\alpha$'s nine subspaces of the six  experiments are observed together, we obtain 54 samples. These samples can still be interpreted by $\sigma_\alpha$ with significance (Spearman's $\rho$ = 0.3554, $p$(2-tailed) = 0.00836, $N$=54).
        \item Similarly, with the $\sigma_\beta$'s six subspaces under the six experiments, we gain 36 samples, but the result is weakly significant (Spearman's $\rho$ = 0.31767, $p$(2-tailed) = 0.05903, $N$=36).
    \end{itemize}
\end{itemize}
 \begin{table}
\caption{Ordinary least squares results of experimental $L_{P11}$ based on the eigencycle set}
 \vspace{-0.2cm}
 \begin{center}
\begin{tabular}{crrrrrc}
 \toprule
\tabincell{c}{eigencycle\\ set}&coef.&\tabincell{c}{coef. \\ $t$}&\tabincell{c}{coef.\\ $p>|t|$}& \tabincell{c}{cons \\ $t$}	& \tabincell{c}{cons \\$p>|t|$} & \tabincell{c}{cons \\ conf.~interval}\\
 	\hline
$\sigma_{.8i}$& 0.057 &27.36& 0.000 &0.11 & 0.910 & [$-$0.001 ~~0.001]\\
\hline
$\sigma_\alpha$& 0.006 &4.43 & 0.003 &-10.94& 0.000&[$-0.006$ $-0.004$] \\
\hline
$\sigma_\beta$ &$ 0.005 $ & $3.31$ &0.030&$-0.77$& 0.487&[$-0.002$ ~~0.001] \\ \hline
\end{tabular}
\end{center}
\vspace{-0.2cm}
\footnotesize{
Note:
By the software Stata, version SE 15.1.
}
\label{tab:P11_alpha_beta}
\end{table}

\begin{flushleft}
\textbf{Interpretation of result 2}
\end{flushleft}
\begin{itemize}
\item \textbf{Finding the hyperfine structure~------~}
$L_{P11}$ and $\sigma_{\alpha}$ show significant correlation, indicating that $\sigma_{\alpha}$
can explain the hyperfine structure in the nine subspaces $\Omega^{(26,27,28,36,37,38,46,47,48)}$.
Beyond $\sigma_{.8i}$, which predicts that the nine points have the same value,
$\sigma_{\alpha}$ can also capture the differences among them.
At the same time, the response mode reflected by $\sigma_\alpha$
is consistent with the best response mode in behavior game theory,
which is also consistent with the win-stay-lose-shift mode in human decision-making \cite{Behavioral2003}.
\item \textbf{Puzzle on $\sigma_\beta$'s result~---~}
$L_{P11}$ and $\sigma_{\beta}$ have significant positive correlation, which cannot be explained by behavioral game theory. We suggest that
this is the geometry factor effect,
which is a natural result of
the eight-dimensional presentation of the dynamics equation \ref{eq:repliequl}.
\end{itemize}

\subsection{Eigencycle spectrum analysis \label{R4}}
\begin{flushleft}
\textbf{Descript of result 3:}
\end{flushleft}

With the eigencycle set as the orthogonal basis, we can explore high-dimensional dynamics through spectrum analysis in the form of
Eq. (\ref{eq:eigdeco1}).
From the two subsections above, we can conclude that $\sigma_{.8i}$, $\sigma_{\alpha}$, and $\sigma_{\beta}$ simultaneously affect $L_{P11}$.

%
%
%
%

\begin{flushleft}
\textbf{Supporting data of result 3:}
\end{flushleft}

\begin{table}
\caption{Multiple linear regression of experimental $L_{P11}$ by
$\sigma_{.8i}$, $\sigma_{\alpha}$, and $\sigma_{\beta}$}
 \vspace{-0.2cm}
\begin{center}
\begin{tabular}{ c|r r r r r }
 \toprule
 \vspace{0.1cm}
$L_{P11}$&	Coef.~~~~&	Std. Err.~~&	$t$~~~&	$P\!>\!|t|$&~~~~[95\% Conf.	Interval]~~~~~\\\hline
$\sigma_{.8i}$		&	~.0565190	&	~.0015589	&	36.25	&	~.000	&	 ~.0533014~~	~.0597365 ~~	\\	
$\sigma_{\alpha}$		&	~.0057181	&	~.0015589	&	3.67	&	~.001	& ~.0025006~~	~.0089356 ~~	\\	
$\sigma_{\beta}$	&	~.0045221	&	~.0015778	&	2.87	&	~.009	&	 ~.0012656~~	~.0077785 ~~	\\
cons		&	$-$.0000876	&	~.0002982	&	$-$0.29	&	~.772	&	 $-$.000703~~	~.0005279 ~~	\\	\hline
\end{tabular}
\end{center}
\label{tab:multi_reg}
\end{table}

Multiple regression analysis is a powerful technique used for predicting the unknown value of a variable from the known value of two or more variables-  also called the predictors.
Table \ref{tab:multi_reg} shows the supporting data. Multiple linear regression helps us interpret the experimental observation based on three theoretical eigencycle sets:

\begin{equation}\label{eq:empricalfor}
 \underset{SE}{L_{P11}}\underset{=}{=}\underset{(0.002)}{0.057~\!\sigma_{.8i}}+\underset{(0.002)}{0.006\sigma_{\alpha}}+\underset{(0.002)}{0.005\sigma_{\beta}}-0.00009,
\end{equation}
Function~\ref{eq:empricalfor} is evidence that $\sigma_{.8i}$, $\sigma_{\alpha}$, and $\sigma_{\beta}$ have a significantly positive effect on $L_{P11}$. $0.057$ is the partial regression coefficient of $\sigma_{.8i}$, which shows that, with the influence of $\sigma_{\alpha}$ and $\sigma_{\beta}$ held constant, a one-unit increase in $\sigma_{.8i}$ increases $L_{P11}$ by 0.057 units.
Similar significance from $\alpha$ and $\beta$ can be obtained.

\begin{flushleft}
\textbf{ Result 3's interpretation:}
\end{flushleft}
\begin{itemize}
\item
Consistent with the two subsections above, the results in this subsection are also significant. That is,
(1) The experimental observation is mainly determined by $\sigma_{.8i}$.
(2) Compared with $\sigma_{.8i}$,
$\sigma_\alpha$ and $\sigma_\beta$ have a weaker but significant effect.
 \item
 Because of constant cannot reject 0 (linear regression, $t=-0.29$, $p=0.772$), $\sigma_{.8i}$, $\sigma_\alpha$ and $\sigma_\beta$ are ideal basic vector for spectral analysis.
\item
The eigencycle set can be an ideal basis for spectrum analysis,
which is helpful for revealing the motion characteristics of high-dimensional game dynamics.
The results in Table \ref{tab:reg8i4i14i2}are a significant illustration of this fact.
\end{itemize}

\section{Discussion and conclusion}
\begin{itemize}
\item Contributions. This study is unique in its discovery of
        the fine structure, the cycles in high dimension game,
        in the long-existed human behavior O'Neill game experiments.
        To our knowledge, it is the first report of the high dimensional cycle.
        We also find evidence on the existence of the hyperfine structure in the human cycling spectrum (see Figure \ref{fig:fine_structure}).
        Meanwhile we have developed the eigencycle set decomposition approach.
        Through the invariant between the eigenvector components
        in a periodic dynamics system, by which we can
        test out the fine structure spectrum of human social cycling.
\item On the implications of the discovery,
we suggest that it may deeply influence our view on human game behavior.
The conflict between the complex in experiment
and the simply in theory seems visible.
(1) The complex of the experiments game behaviors,
which is high stochastic, discrete time, discrete strategy, non-linear, and high dimensional system, short time series, a extensive studied data which has existed more than 3 decades;
(2) The simply of the theoretical predictors, which is differential, continue time, continue strategy space, and solvable by the tools from the standard concepts of the dynamics system theory. However, referring to the finding of the fine dynamics structure, they are compatible.
It is well known that, the concepts and tools (e.g., stationary concept, Jacobian,  eigensystem concept, Near periodic orbits, Poincare Section, Centre manifold theory) in the dynamics system theory are powerful. Now we see that,
the reality and accuracy of the human behaviour game dynamics, then  the evolutionary game theory, can be well improved along the standard dynamics system theory.
%
%
\item   On the implication of the eigencycle approach, we suggest that, it may be an improvement to the standard dynamics system theory.
Not only can bring game dynamics system theory
to experimental game dynamics systematically,
it can be part of the standard analytical approach to where dynamics systems theory can be applied.
This approach is bases on the invariant between two components of a given eigenvector, which, to our knowledge, is novel.
Considering the widely application of the dynamics system theory,
we hope to have more strictly mathematics study
and potential empirical verification on the validation of the approach.
Some details on the comparison of our approach with
existed approach are shown in appendix
section \ref{app:rw_dst}.
\item  We concern the questions raised during this study. Our finding have demonstrated the significant existence of cycles in high dimension game by the eigencycle approach.
Although out of the scope of this study, there are some enigma remained,

\begin{itemize}
    \item The coefficients in Eq.~(\ref{eq:empricalfor}) are not quantitatively explained in a clear manner. These coefficients may have physical meaning, such as the distribution of electrons in their orbit (eigenstates) obeying the Boltzmann distribution. But what governors the coefficients is unclear. On this question,
    some simulation results have shown that
    the coefficients is strongly depends
    on individual decision making methods \cite{2021Qinmei},
    but the answer remains fuzzy.
    \item The eigencycle approach offers new opportunities for formal
    dynamics model analysis, but also has its own limitations.
     First, it is based on linear approximation which is common in eigen system analysis. So, the application
     on large deviation condition needs caution.
     Second,  we ignore the phase interfering between two components from two eigenvectors, but if in a delegation imaginary part of eigenvalue condition (although not common),
     the long-time and sufficient noise condition is necessary (see the proof in appendix section \ref{app:itf_diff_ev}). So, for further strictly application, the boundary of the eigencycle approach needs further investigate.
\end{itemize}

\item  Our final concern is the central concept of game dynamics theory, or evolutionary game theory. Classical game theory has its central concept, Nash equilibrium, which reality, accuracy and applications have been well established. However, game dynamics theory has not such well established central concept.
The ESS (evolution stable strategy) concept,  which is mainly qualitative concept. Since recent experiments quantitative results accumulated \cite{Binmore2001Minimax,dan2014,wang2014,dan2016book},
a quantitative concept is more urgent.
The fixed point theorem leads to the Nash equilibrium,
which has become the central concept
of the classical game statics theory.
Have the finding the high dimensional game eigencycles, 
we calculate the 'invariant manifolds' and compare with eigencycles ant net transit. 
See appendix section \ref{app:im_ec} and \ref{app:net_trans} for details. 
As straightly impress, we suggest that  
'invariant manifolds' could be a candidate as the central concept,
which be used to capture complex dynamics and the long term behavior pattern of game dynamics.
This concept, not only have solid mathematics background, 
but also can be verified in laboratory experiment. 
In the field of game dynamics theory and experiment, on dynamics pattern, there exists a common question \cite{dan2020}. 
\begin{center}
\textit{In the background is a metaphysical question: what is a cycle?}
\end{center}
By the fine structure out of the eigencycles in O'Neill games data, we wish the 'invariant manifolds' can be a concrete answer. 
\end{itemize}

~\\
\textbf{Acknowledgement}: We thank O'Neill, Ken Binmore,
and Yoshitaka Okano for providing the data.
Thanks to Yijia Wang for her critical suggestion for Zhijian
to obtain the pair of orthogonal basis ($\alpha, \beta$).
The first author could be changed from Wang to Yao at any time
during the revising and publishing processes of the top five economics journal.
Wang designed the approach and Yao discovered the fine structure.

\bibliographystyle{plain}
\bibliography{Sec6referen2}

\newpage
\section{Appendix}
\subsection{Appendix 1: Geometrical interpretation of eigencycle}\label{app:geo_eigc}

Geometrical interpretation of eigencycle depend on the
geometrical interpretation of complex eigenvectors in
a system of differential equations.

A key element in dynamics system theory
(modern control theory or linear systems theory) is the
notion of the system eigenvalues, i.e., the eigenvalues
of the Jacobian matrix. For simplicity, we restrict
ourselves to the endogenous dynamics of the system.
Suppose that an initial probability distribution can be expressed as
a linear combination of the eigenvectors $\xi$ as \cite{Lu5083},
\begin{equation}\label{eq:eigdeco1}
 {x}(0) =  c_1 {\xi}_1 + c_2 {\xi}_2 + ... + c_n {\xi}_n,
\end{equation}
wherein $\xi_i$ is associated eigenvector
of the eigenvalue  $\lambda_i$;
The probability will evolve in time according to
\begin{equation}\label{eq:eigdeco2}
{x}(t) = e^{\lambda_1t} c_1 {\xi}_1 + e^{\lambda_2 t} c_2 {\xi}_2
         + ... + e^{\lambda_n t} c_n {\xi}_n,
\end{equation}
Then the $i$ component of $x$ is
\begin{equation}\label{eq:x_i_t}
x_i(t) = \sum_{j=1}^n c_{j} \cdot \eta_{ij}
         \cdot \exp(\lambda_j t) ~~~~~ \{i,j\} \in \{1,2,.... n\}.
\end{equation}
For a given $\lambda_j$, the $c_j$ is invariant
for any  components of the eigenvector $\xi_j$.
Such that a given pair of the components will evolve respectively as
\begin{eqnarray}
\left\{ \begin{array}{rcl}
    x_m(t) &=& c_j \cdot || \eta_m || \exp(\arg(\eta_m)) \cdot\exp(\lambda_j t)  \\
    x_n(t) &=& c_j \cdot || \eta_n || \exp(\arg(\eta_n)) \cdot\exp(\lambda_j t)
     \end{array}
     \right.
\end{eqnarray}
Now we can show the geometric pattern of
the evolution of $x_m$ and $x_n$.

Suppose that the $\lambda_i$ is pure  imaginary number,
in the two dimension (m,n) phase space,
the curve is (1,1)-Lissajous,
which is closed and periodic curve;

Meanwhile, the signed area is $\sigma^{(mn)}$ factor
by $c_j^2$ as
\begin{equation}
      c_j^2 \sigma^{(mn)}=\pi \cdot ||c_j||^2 \cdot ||\eta_m||
      \cdot ||\eta_n|| \cdot  \sin\left(\arg(\eta_m)-\arg(\eta_n)\right),
\end{equation}
Important is that for a given initial condition,
the coefficients $c_j$ is invariant along time.
So, in all of the two dimensional subspace,
the signed area is multiple by the constant $||c_j||^2$.
That is
\begin{eqnarray}
  \frac{\sigma^{(mn)}}{\sigma^{(m'n')}} &=&
  \frac{||\eta_m|| \cdot ||\eta_n|| \cdot  \sin\left(\arg(\eta_m)-\arg(\eta_n)\right)}
  {||\eta_m'|| \cdot ||\eta_n'|| \cdot  \sin\left(\arg(\eta_m')-\arg(\eta_n')\right)} \\
  &=& \frac{\Re(\eta_m)\Im(\eta_n) - \Re(\eta_n)\Im(\eta_m) }
     {\Re(\eta_{m'})\Im(\eta_{n'}) - \Re(\eta_{n'})\Im(\eta_{m'})}
\end{eqnarray}
Then the rate between the signed area is independent of
$c_j$ or $\lambda_j$ in Eq.~(\ref{eq:x_i_t}).

\newpage
\subsection{Appendix 2: Invariant between eigencycle and
the angular momentum}\label{app:sigma_L}

The invariant between
the eigencycle and the angular momentum is important in this study.
Because, only having this invariant, we can using eigencycle as predictor
for experimental angualr momentum.
Here, the eigencycle as predictor can be obtained
by the eigenvector from dynamics equation, and
the angular momentum can he obtained in time series.

In this section, we show that,  the rate between
the value of the eigencycle angular momentum $L_{mn}$
and the value of the eigencycle  $\sigma_{mn}$ is \textbf{invariant}
in respect to the subspace $\Omega^{mn}$, in which $m,n \in \{1,2,...,N\}$.

Main idea to prove the invariant is
to calculate the angular momentum of the eigencycle explicitly.

\subsubsection{Interfere of two components from the same eigenvector}
First we need to describe the state $x(t)$ explicitly. Suppose that, in Eq. \ref{eq:eigdeco2}, the eigenvalue $\lambda_i$ is not pure imaginary number,
but a complex number $\lambda = \Re(\lambda) + i \Im({\lambda})$, wherein $\Re(\lambda) \neq 0, \Im({\lambda}) \neq 0$. Then we have the $m$-th component
$$x_m(t) = c \cdot \exp(t \cdot (\Re(\lambda) + \Im(\lambda) \cdot i)) \cdot (\Re(\eta_m) + \Im(\eta_m) \cdot i) $$
herein we ignore the subscribe of $c$ and $\lambda$.
The real part of $ x_m(t) $
\begin{eqnarray}
x^r_m(t)&=&~~c \cdot \Re(\eta_m) \cdot \exp(\Re(\lambda) \cdot t) \cdot \cos(\Im(\lambda) \cdot t) \\
\nonumber& & - c \cdot \Im(\eta_m) \cdot \exp(\Re(\lambda) \cdot t) \cdot \sin(\Im(\lambda) \cdot t)
\end{eqnarray}

Then the velocity of the state can be expresses as

\begin{eqnarray}
\frac{d}{dt} x^r_m(t)&=&  c \cdot (\Re(\eta_m) \cdot \Re(\lambda)- \Im(\eta_m) \cdot \Im(\lambda)) \cdot \exp(\Re(\lambda) \cdot t) \cdot \cos(\Im(\lambda) \cdot t) \\
 \nonumber&& - c \cdot (\Im(\eta_m) \cdot \Re(\lambda) - \Im(\lambda) \cdot \Re(\eta_m)) \cdot  \exp(\Re(\lambda) \cdot t) \cdot \sin(\Im(\lambda) \cdot t)
\end{eqnarray}

Such, in (m,n) subspace, the instantaneous signed area change
(the angular momentum) is
    \begin{eqnarray}
        L_{mn}(t)&= & x_{mn}^r(t) \times \frac{d}{dt} x_{mn}^r(t) \\
            &=& [x_{m}^r(t),~ x_{n}^r(t)]\times [\frac{d}{dt} x_{m}^r(t),~ \frac{d}{dt} x_{n}^r(t)]^T \\
            &=&x^r_m(t) \cdot \frac{d}{dt} x^r_n(t) - x^r_n(t) \cdot \frac{d}{dt} x^r_m(t) \\
            &=&c^2 \cdot \Im(\lambda) \cdot \exp(2 \cdot \Re(\lambda) \cdot t) \cdot (\Im(\eta_m) \cdot \Re(\eta_n) - \Im(\eta_n) \cdot \Re(\eta_m))
    \end{eqnarray}
So that, for any two sub-spaces $\Omega^{(mn)}$ and $\Omega^{(m'n')}$,
we have the ratio of the the angular momentum at $t$ as
\begin{eqnarray}
R_{(mn,m'n')}(t) &=& \frac{L_{mn}(t)}{L_{m'n'}(t)} \\
    &=& \frac{ c^2\, \Im{(\lambda)}\, \mathrm{\exp(2\, \Re{(\lambda)}\, t)}\,  \cdot  \left(\Im{(\eta_m)}\, \Re{(\eta_n)} - \Im{(\eta_n)}\, \Re{(\eta_m)}\right)}
    { c^2\, \Im{(\lambda)}\, \mathrm{\exp}(2\, \Re{(\lambda)}\, t)\,  \cdot   \left(\Im{(\eta_m')}\, \Re{(\eta_n')} - \Im{(\eta_n')}\, \Re{(\eta_m')}\right)} \\
    &=& \frac{\Re{(\eta_m)}\,\Im{(\eta_n)} - \Im{(\eta_m)}\, \Re{(\eta_n)} } { \Re{(\eta_m')}\,\Im{(\eta_n')} - \Im{(\eta_m')}\, \Re{(\eta_n')}} \\
    &=&   \frac{\sigma^{(mn)}}{\sigma^{(m'n')}}
\end{eqnarray}
So, we having following results.
~~\\
\fbox{\parbox{\textwidth}{
 \textbf{Proposition 1} \\
 For a given eigenvector $\xi$, in all of the two dimension sub-spaces $\Omega^{(m,n)}$
 at any given $t$, the rate between the instantaneous angular momentum value and eigencycle value is equal,  \\
\begin{equation}
    \frac{L_{mn}(t)}{\sigma^{(mn)}} = \textbf{C}(t) ~~~~ \forall (m,n) \in (1,2,...,N)
\end{equation}
Such that, we can expect that, for various subspace $(m,n)$,
the average accumulated angular momentum
$L_{mn}$ divided by its $\sigma^{(mn)}$ is  \textbf{invariant},
\begin{equation}
   \frac{1}{t_1 - t_0} \sum_{t=t_0}^{t_1}\frac{L_{mn}(t)}{\sigma^{(mn)}}
   = \frac{1}{t_1 - t_0} \sum_{t=t_0}^{t_1}\frac{L_{m'n'}(t)}{\sigma^{(m'n')}}
   = \textbf{C}  ~~~~ \forall (m,n;m',n') \in (1,2,...,N)
\end{equation}
   So, by take $\sigma^{(mn)}$ as predictor (independent variable) and take $L_{mn}$ as the observation (dependent variables), the a linear regression line, with truncation term being 0,
   is expected.
}}

\subsubsection{Interfere of two components from two different eigenvector}
\label{app:itf_diff_ev}

 For two given eigenvector $\xi_1$ and $\xi_2$, in all of the two dimension sub-spaces the time average of the $L_{mn'} = 0$ in which
 $ m\in \xi_1 \cap n' \in \xi_2$.  \\
 \\

 Denote the eigenvalue as $\lambda$,
 and $\Re (\lambda) :=\delta$, $\Im (\lambda) :=\omega$,
 the initial phase as $\phi$, the instantaneous angular momentum at $t$ can be expressed as,
\begin{eqnarray}
L_{(m,n')} (t)
     &=& x \times \frac{dx}{dt}  \\
     &=& \Im \Big (\Big[ c_1 \exp( (\delta_1 + \,i \, \omega_1)\, t) A_m \exp( \,i \,\phi_m)\Big]^\dag
     \cdot \Big[c_2 \exp( (\delta_2 +\,i \,\omega_2)  t) A_{n'} \exp(\,i \, \phi_{n'})\Big] \Big ) \\
     &=& \Im \Big (\Big[ c_1 \exp( (\delta_1 - \,i \, \omega_1)\, t) A_m \exp(-\,i \,\phi_m)\Big]
     \cdot \Big[c_2 \exp( (\delta_2 +\,i \,\omega_2)  t) A_{n'} \exp(\,i \, \phi_{n'})\Big] \Big ) \\
     &=&  c_1  c_2 A_m A_{n'} \exp\Big((\delta_1+\delta_2)\, t\Big) \, \Im \Big (\exp(\,i \,
     ((- \omega_1 + \omega_2)\,t - \phi_m + \phi_{n'}))\Big ) \\
     &=&  c_1  c_2 A_m A_{n'} \exp\Big((\delta_1+\delta_2)\, t\Big) \, \sin \Big (
     (- \omega_1 + \omega_2)\, t - \phi_m + \phi_{n'}\Big )
\end{eqnarray}
We set the term $c_1  c_2 \exp\Big((\delta_1+\delta_2)\, t\Big) = 1$,
which does not related to angular motion strictly. Then we have
\begin{enumerate}
\item \textbf{Two inner components' interfere} \\
    When  $m \in \xi_1$ and $n' \in \xi_1$ simultaneously, it is two inner components' interfere. As belong to same eigenvector and having same eigenvalue, then $\omega_1 = \omega_2$.
    Then the long time average of $L$ is,
    \begin{eqnarray}
    \overline{L_{(m,n')}} &=& \lim_{T \rightarrow \infty}
                   \frac{1}{T}\int_0^T A_m A_{n'}
                   \sin \Big (- \phi_m + \phi_{n'}\Big )
                   \, dt \\
                         &=& \frac{1}{\pi}\sigma^{(mn')}
    \end{eqnarray}
    is not 0. One can see that,
    the eigencycle value shown in main text can be re-archived,
    only less the constant $1/\pi$.
\item \textbf{Two cross components' contribution in no delegation condition} \\
    When $\xi_1 \neq \xi_2$ and $\omega_1 \neq \omega_2$,
    the long time average,
    \begin{eqnarray}
    \overline{L_{(m,n')}}
       &=& \lim_{T \rightarrow \infty}
           \frac{1}{T}\int_0^T A_m A_{n'}
           \sin \Big ((- \omega_1 + \omega_2)\, t
           - \phi_m + \phi_{n'}\Big ) \\
        && + A_{m'} A_{n}
           \sin \Big ((- \omega_2 + \omega_1)\, t
           - \phi_{m'} + \phi_{n}\Big ) \, dt \\
       &=& 0
    \end{eqnarray}
\item \textbf{Two cross components'  contribution in delegation condition without noise} \\
    When $\xi_1 \neq \xi_2$ and $\omega_1 = \omega_2$,
    this is the delegation condition.
    when facing $\lambda = 0.4i$ condition.
    In such condition, we can not strictly prove that
    the time average of the cross components contribution
    is zero.
        \begin{eqnarray}
    \overline{L_{(m,n')}}
       &=& \lim_{T \rightarrow \infty}
           \frac{1}{T}\int_0^T A_m A_{n'}
           \sin \Big ((- \omega_\alpha + \omega_\beta)\, t
           - \phi_m + \phi_{n'}\Big ) \\
        && + A_{m'} A_{n}
           \sin \Big ((- \omega_\beta + \omega_\alpha)\, t
           - \phi_{m'} + \phi_{n}\Big ) \, dt \\
       &=& \lim_{T \rightarrow \infty}
           \frac{1}{T}\int_0^T A_m A_{n'}
           \sin \Big ( - \phi_m + \phi_{n'}\Big ) \\
        && + A_{m'} A_{n}
           \sin \Big ( - \phi_{m'} + \phi_{n}\Big ) \, dt \\
       &\neq& 0.
    \end{eqnarray}
    Because, in general case, the two equality,
    $A_{m'} A_{n} = A_m A_{n'}$ and
    $- \phi_m + \phi_{n'}  = \phi_{m'} - \phi_{n}$,  is not fulfilled.
\item \textbf{Two  components' cross contribution in delegation condition with noise}\\
    So called stochastic condition,
    we assume the white noise interrupts the system,
    and the system is rebuild, at this moment,
    the coefficient $c_i$ and the initial phase $\phi_i$
    changes randomly.
    As the eigen value and eigen vector remain unchanged,
    the system can be expressed as,
    $$p(t) = \sum_i \, c_i  \, \xi_i \exp(\lambda_i \,t + \theta_i).$$

    \begin{eqnarray}\label{eq:dele_noise}
    \overline{L_{(m,n')}} &=& \lim_{K \rightarrow \infty}
                   \frac{1}{\sum_{k=0}^{K}\tau_k} \sum_{k=0}^{K}\int_0^{\tau_k} A_m A_{n'}
                   \sin \Big (\!-\! (\phi_m + \theta_{1k}) + (\phi_{n'}+ \theta_{2k})\Big )
                   \, d\,t  \\
                         &=& 0
    \end{eqnarray}
    Herein, $\theta_{1k}$ the initial phase angle of $\xi_1$
    after the $k$-th shock of noise.
    Following condition is necessary for this result hold:
    \begin{enumerate}
      \item White noise, and after a random interrupt, the $\theta_1$ and $\theta_2$ distribute between [$-\pi, \pi$] equally.
      \item Sufficient noise shack times limit, or saying there are sufficient times of the interrupt, that is $K \rightarrow \infty$.
      \item Long time limit, there are sufficient long time  of the time series, that is $\sum_k \tau_k \rightarrow \infty$.
    \end{enumerate}
In summery, the 0 comes out of the symmetry of the initial phase.
In sufficient longtime, each $\theta_{1k}-\theta_{2k}$ has its oppose condition, then the average of the sinusoidal term is 0.
In section \ref{app:im_ec} with a numerical verification approach,
we provide an explanation on the results shown in Eq. (\ref{eq:dele_noise}).
\end{enumerate}

\fbox{\parbox{\textwidth}{
 \textbf{Proposition 2} \\
 Interfere of two components from two different eigenvector is vanish,
 when following condition hold:\\
    \begin{enumerate}
      \item White noise condition: after a random interrupt, the initial phase
            of the two eigenvector distribute between [$-\pi, \pi$] equally.
      \item Sufficient noise shock: there are sufficient times of
            the interrupt, that is $k \rightarrow \infty$.
      \item Long time limit: there are sufficient long time  of the time series.
    \end{enumerate}
}}

\subsection{Appendix 4: The eigen system derivation and symmetry analysis}\label{app:derive_eig}
 \subsubsection{The eigen system derivation}

For the given payoff matrix, assume $x_i$ is the proposition of player 1 play strategy-$i$, the expected payoff for each strategy is,
$$ \left(\begin{array}{c} U_{x_1}\\U_{x_2}\\U_{x_3}\\U_{x_4}\\U_{y_1}\\U_{y_2}\\U_{y_3}\\U_{y_4}\\ \end{array}\right)= \left(\begin{array}{c} y_{1} - y_{2} - y_{3} - y_{4}\\ y_{3} - y_{2} - y_{1} + y_{4}\\ y_{2} - y_{1} - y_{3} + y_{4}\\ y_{2} - y_{1} + y_{3} - y_{4}\\ x_{2} - x_{1} + x_{3} + x_{4}\\ x_{1} + x_{2} - x_{3} - x_{4}\\ x_{1} - x_{2} + x_{3} - x_{4}\\ x_{1} - x_{2} - x_{3} + x_{4} \end{array}\right)
$$ \\
Herein, $U_{x_i}$ is the payoff of the strategy $x_i$. Then,
the mean payoff $\overline{U_X}$ and $\overline{U_Y}$ for the two populations (players) can be expressed respectively as
\begin{eqnarray}
\overline{U_X} &=&   - x_{1}\, \left(y_{2} - y_{1} + y_{3} + y_{4}\right) - x_{2}\, \left(y_{1}  + y_{2} - y_{3} - y_{4}\right) \nonumber \\
 & & - x_{3}\, \left(y_{1} - y_{2} + y_{3} - y_{4}\right) - x_{4}\, \left(y_{1} - y_{2} - y_{3} + y_{4}\right) \\
\overline{U_Y}  &=&  y_{1}\, \left(x_{2} - x_{1} + x_{3} + x_{4}\right) + y_{2}\, \left(x_{1} + x_{2} - x_{3} - x_{4}\right) \nonumber \\
 & & + y_{3}\, \left(x_{1} - x_{2} + x_{3} - x_{4}\right) + y_{4}\, \left(x_{1} - x_{2} - x_{3} + x_{4}\right)
\end{eqnarray}

Referring to the replicator dynamics, shown in main text as Eq. (\ref{eq:repliequl}), the expanded form is:
\begin{eqnarray}\label{eq:8_repli}
\frac{d}{dt}x_1 &=&  x_{1}\, (y_{1} - y_{2} - y_{3} - y_{4} + x_{1}\, \left(y_{2} - y_{1} + y_{3} + y_{4}\right) + x_{2}\, \left(y_{1} + y_{2} - y_{3} - y_{4}\right)\\ &&
x_{3}\, \left(y_{1} - y_{2} + y_{3} - y_{4}\right) + x_{4}\, \left(y_{1} - y_{2} - y_{3} + y_{4}\right) \nonumber \\
\frac{d}{dt}x_2  &=&  x_{2}\, (y_{3} - y_{2} - y_{1} + y_{4} + x_{1}\, \left(y_{2} - y_{1} + y_{3} + y_{4}\right) + x_{2}\, \left(y_{1} + y_{2} - y_{3} - y_{4}\right) \\ &&
+ x_{3}\, \left(y_{1} - y_{2} + y_{3} - y_{4}\right) + x_{4}\, \left(y_{1} - y_{2} - y_{3} + y_{4}\right) \nonumber \\
\frac{d}{dt}x_3 &=&  x_{3}\, (y_{2} - y_{1} - y_{3} + y_{4} + x_{1}\, \left(y_{2} - y_{1} + y_{3} + y_{4}\right) + x_{2}\, \left(y_{1} + y_{2} - y_{3} - y_{4}\right)\\ &&
 + x_{3}\, \left(y_{1} - y_{2} + y_{3} - y_{4}\right) + x_{4}\, \left(y_{1} - y_{2} - y_{3} + y_{4}\right) \nonumber \\
\frac{d}{dt}x_4 &=&  x_{4}\, (y_{2} - y_{1} + y_{3} - y_{4} + x_{1}\, \left(y_{2} - y_{1} + y_{3} + y_{4}\right) + x_{2}\, \left(y_{1} + y_{2} - y_{3} - y_{4}\right)\\ &&
  + x_{3}\, \left(y_{1} - y_{2} + y_{3} - y_{4}\right) + x_{4}\, \left(y_{1} - y_{2} - y_{3} + y_{4}\right) \nonumber \\
\frac{d}{dt}y_1 &=&  - y_{1}\,(x_{1} - x_{2} - x_{3} - x_{4} + y_{1}\, \left(x_{2} - x_{1} + x_{3} + x_{4}\right) + y_{2}\, \left(x_{1} + x_{2} - x_{3} - x_{4}\right) \\ &&
 + y_{3}\, \left(x_{1} - x_{2} + x_{3} - x_{4}\right) + y_{4}\, \left(x_{1} - x_{2} - x_{3} + x_{4}\right) \nonumber \\
\frac{d}{dt}y_2  &=&  - y_{2}\, (x_{3} - x_{2} - x_{1} + x_{4} + y_{1}\, \left(x_{2} - x_{1} + x_{3} + x_{4}\right) + y_{2}\, \left(x_{1} + x_{2} - x_{3} - x_{4}\right)\\ &&
 + (y_{3}\, \left(x_{1} - x_{2} + x_{3} - x_{4}\right) + y_{4}\, \left(x_{1} - x_{2} - x_{3} + x_{4}\right) \nonumber \\
\frac{d}{dt}y_3  &=&  - y_{3}\, (x_{2} - x_{1} - x_{3} + x_{4} + y_{1}\, \left(x_{2} - x_{1} + x_{3} + x_{4}\right) + y_{2}\, \left(x_{1} + x_{2} - x_{3} - x_{4}\right) \\ &&
 +  y_{3}\, \left(x_{1} - x_{2} + x_{3} - x_{4}\right) + y_{4}\, \left(x_{1} - x_{2} - x_{3} + x_{4}\right) \nonumber \\
\frac{d}{dt}y_4  &=&  - y_{4}\, (x_{2} - x_{1} + x_{3} - x_{4} + y_{1}\, \left(x_{2} - x_{1} + x_{3} + x_{4}\right) + y_{2}\, \left(x_{1} + x_{2} - x_{3} - x_{4}\right)\\ &&
 + \,  y_{3}\, \left(x_{1} - x_{2} + x_{3} - x_{4}\right) + y_{4}\, \left(x_{1} - x_{2} - x_{3} + x_{4}\right) )   \nonumber
\end{eqnarray}

The rest point of the game can be solved with the concentration
$$0 < x_i, y_i < 1, ~~~~ i \in [1,2,3,4]$$
Then the mixed strategy solution $x^*$ can be archive,
$$x^*:=(x_1^*,x_2^*,x_3^*,x_4^*,y_1^*,y_2^*,y_3^*,y_4^*)=(2/5,1/5,1/5, 1/5,2/5,1/5,1/5,1/5).$$

The Taylor series expansion of $dx/dt$ about this point $x^*$ is:
$$\frac{d}{dt}x_i = \frac{d}{dt}x_i \Big |_{x^*} + \dfrac{\partial (\frac{d}{dt}x_i)}{\partial x_j} (x_j - x_j^*) + O(\triangle x^2)$$

We assume $v$ sufficiently smooth and differentiable for the
Taylor expansion to exist.
At a critical point, the first term of the Taylor expansion vanishes
(by definition). We consider only the second term.

At this points, the Jacobian matrix (character matrix)
\begin{eqnarray}
  \mathbf J \Big |_{x^*} &=&  \begin{bmatrix}
                    \dfrac{\partial (\frac{d}{dt}x_1)}{\partial x_1} & \cdots & \dfrac{\partial \frac{d}{dt}(x_1)}{\partial x_8}\\
                    \vdots & \ddots & \vdots\\
                    \dfrac{\partial \frac{d}{dt}(x_8)}{\partial x_1} & \cdots & \dfrac{\partial \frac{d}{dt}(x_8)}{\partial x_8} \end{bmatrix}_{ {|x^*}}  \\
  &=& \frac{1}{25}\begin{bmatrix}	\begin{array}{rrrrrrrr}
	 2 & 2 & 2 & 2 & 12 & -8 & -8 & -8 \\
	 1 & 1 & 1 & 1 & -4 & -4 & 6 & 6 \\
	 1 & 1 & 1 & 1 & -4 & 6 & -4 & 6 \\
	 1 & 1 & 1 & 1 & -4 & 6 & 6 & -4 \\
	 -12 & 8 & 8 & 8 & -2 & -2 & -2 & -2 \\
	 4 & 4 & -6 & -6 & -1 & -1 & -1 & -1 \\
	 4 & -6 & 4 & -6 & -1 & -1 & -1 & -1 \\
	 4 & -6 & -6 & 4 & -1 & -1 & -1 & -1 \\
\end{array} \end{bmatrix}
\end{eqnarray}

	Using the Characteristic polynomial,
	this character matrix is diagonalizable.
    From which, the eigen valueand eigenvector set can be archived explicitly.
	$https://matrixcalc.org/en/vectors.html$

\subsubsection{Eigen system and invariant manifold}

The suggestion on the invariant manifold can be a candidate
of the central concept is based on the bridge between the
Eigen system and invariant manifold,
both of which are concepts in dynamical systems theory. \cite{2019Roussel}

Eigen system indicates  the eigenvalues and eigenvectors,
and their related concepts, like eigenspace and eigendirection.
An invariant manifold is a topological manifold that is invariant under the action of the dynamical system.

Around the rest point, the eigenvalues and eigenvectors
of the matrix determine the local behavior $dx/dt$. Positive
eigenvalues indicate that $dx/dt$ is directed away from the critical
point (a repelling eigendirection) and negative values the
opposite (an attracting eigendirection).
A complex conjugant pair of eigenvalues indicate that
$dx/dt$ spirals in or out,
depending on the sign of the real part of the eigenvalues.
Thus the linear approximation of $dx/dt$ near the rest point
is characterized by the eigenvalues and eigenvectors.

Application of the eigenvalues and eigenvectors is dramatically important.
\begin{itemize}
  \item The eigenvalues may be used to classify the type of the rest point.
  In the textbook for game dynamics \cite{2011Sandholm}, Bill Sandholm
  has emphasised the stable manifold theorem,
  which relay on the real part of the eigenvalues.
  The stable manifold theorem says that within some
  neighborhood of $x^*$, there is $k$ dimensional local stable manifold
  on which solutions converge to $x^*$ at
  an exponential rate, and an $n-k$ dimensional
  local unstable manifold on which solutions converge to $x^*$
  at an exponential rate if time is run backward.
  \item The eigenvectors may be used to find its invariant manifolds.
  Based
on the linearized stability analysis, the eigenspaces (subspaces of the full phase space spanned by eigenvectors) can be classified as
   \begin{itemize}
   \item   The stable eigenspace, which is the space spanned by the eigenvectors whose corresponding eigenvalues have negative real parts.
   \item   The unstable eigenspace, which is the space spanned by the eigenvectors whose corresponding eigenvalues have positive real parts.
   \item   The centre eigenspace, is the space spanned by the eigenvectors whose corresponding eigenvalues have a real part of zero.
   \end{itemize}
\end{itemize}

Along such classification \cite{2019Roussel}, the O'Neill game, except the real part,
the eigenvectors forms the 6 dimension \textbf{centre eigenspace}.

In this view, the 28 subspaces are
the projection of the invariant manifold
in the centre eigenspace,
which are of the \textbf{central manifold}.

 \subsubsection{The Lissajous figure and its statistics meaning}
	Lissajous figure is the pattern produced by
    the intersection of two sinusoidal curves,
    the axes of which are at right angles to each other.
    In our study case, any two components from an eigenvector
    will perform as 1:1 Lissajous figure.
    In general condition,
    the Lissajous figures are of elliptical shape in the subspace.
    The appearance of the figure is highly sensitive to
    the amplitudes and the phase difference.

    The shape of the curve serving to identify the characteristics of the signals.
    The shape is helpful to identify the relationship of the two signals.
    Its statistics meaning in dynamics processes, as typical examples,
    are dependent on the phase difference between the two sinusoidal signals:
    \begin{itemize}
      \item  $\phi_m-\phi_n = 0$, the Lissajous figure is
             a straight line with an inclination with positive x-axis,
             the two variable is positive depend.
      \item  $\phi_m-\phi_n = \pi$, the Lissajous figure is
             a straight line with an inclination with negative x-axis,
             the two variable is negative depend.
      \item  $\phi_m-\phi_n = \pi/2$, Counter clockwise cycle,
            $m$ is causal of $n$, statistically,
            when $m$ appears then $n$ appears.
      \item  $\phi_m-\phi_n = -\pi/2$, the Lissajous figure is
             in circular shape, clockwise cycle,
             $n$ is causal of $m$, statistically, when $n$ appears then $m$ appears.
    \end{itemize}

 \subsubsection{The eigen system symmetry}
     The symmetry of the game bases on the Jacobian (character matrix) of the game.
     The result is significant after investigating the eigenvectors.
    \begin{description}
    \item[Real eigenvalues and their eigenvectors] 	Notice that, the sum of the columns is
    	$$(1/5, 1/5, 1/5, 1/5, -1/5, -1/5, -1/5, -1/5)$$ which indicates
    	the game value for player 1 is $1/5$ and player 2 is $-1/5$.
    	These two values are the eigen value of the Jacobian;
    	Meanwhile, their related eigenvectors
    	indicates the Nash distribution \cite{2011Sandholm}.
    \item[Image eigenvalues and their complex eigenvectors]
        There are three pairs of image eigenvalue. Their related eigenvectors involves in the cross interactions between the two populations.

    \begin{itemize}
    \item in 1-st ans 5-th dimension, only $.8i$ eigenvector involves.
      The two $.4i$ eigenvector have no consequence on
      the subspace  $\Omega^{(mn)}$ when $m$ or $n$ includes 1 or 5;
      Because, the $.4i$ related eigenvector components of are 0 in 1 and 5.
    \item The cycles of the crossover of the 2-3-4 and 6-7-8 dimensions.
    \item The cycles of the inner 2-3-4 or of the inner 6-7-8 dimensions.
    \end{itemize}

\begin{figure}[h!]
\centering
\includegraphics[scale=0.55]{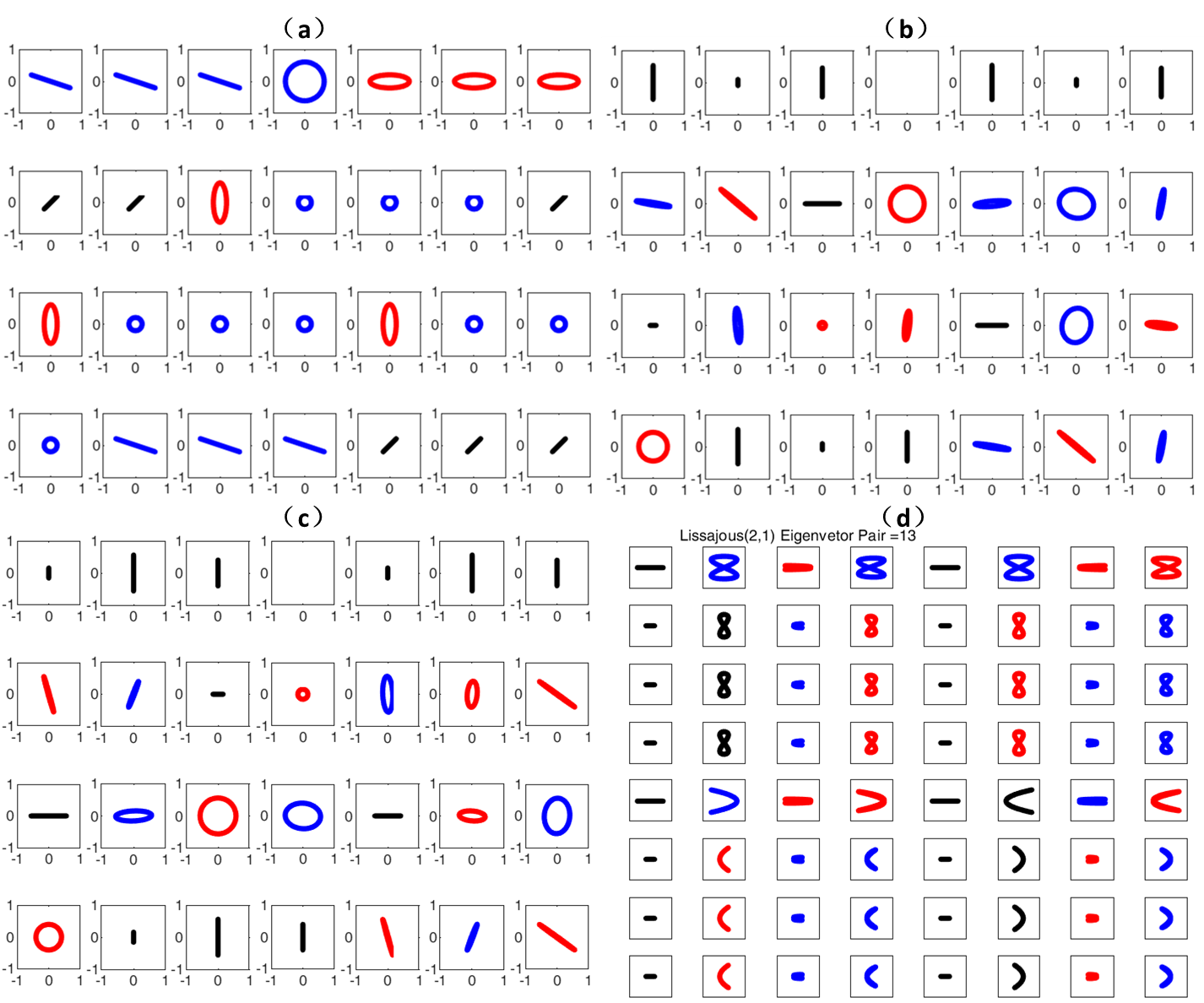}
\caption{Eigencycles in the 28 subspaces from the 3 image eigenvectors
(a) $0.8i$ (b) $0.4i_1$ (c) $0.4i_2$ (d) an illustration of 2:1-lissajous figure. }
\label{fig:improve}
\end{figure}

\end{description}

 \subsection{Appendix 6: Verification of the invariant manifold and eigencycle}\label{app:im_ec}

    Along the concept of the invariant manifold, we can use
    the dynamics system equations set, shown in  Eq. (\ref{eq:8_repli}),
    to calculate the solution for approximate invariant manifold.
    The algorithm in Matlab language is shown following.

\begin{figure}[h!]
    \centering
    \includegraphics[scale=0.75]{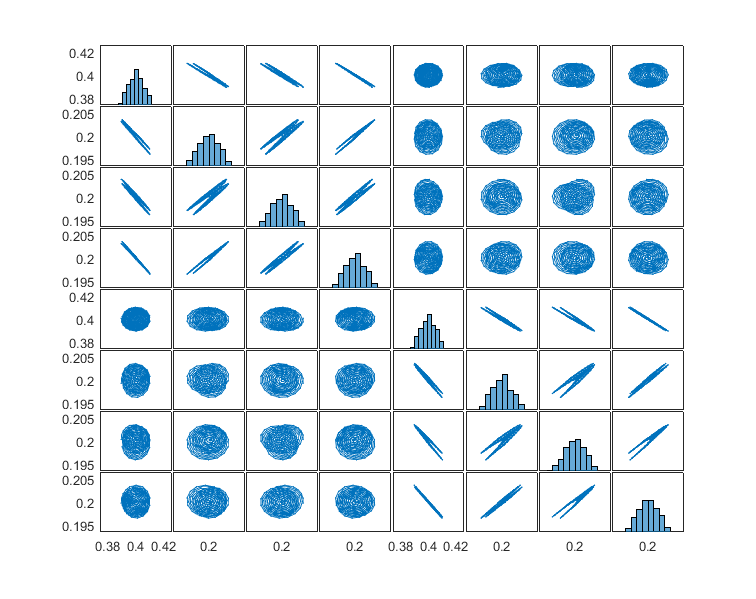}
    \caption{Projection of the 8 dimension
    approximate invariant manifold trajectory
    on the 28 sub-spaces. This is the matrix plot,
    each subplot presents an two variable ($m,n$) interfere.
    The horizon-axis indicates $m$-dimension from left to right is 1 to 8,
    and vertical-axis indicates $n$-dimension from up to down is 1 to 8.}
    \label{fig:ode_invariant_manifold}
    \end{figure}

    As the results, the  invariant manifold
    are approximately shown in figure \ref{fig:ode_invariant_manifold}.
    Obviously, the patterns support
    our abstract of the eigen cycle.
    Several visible results are: \\
    \begin{enumerate}
    \item The cycle or ellipse shape exactly similar to the Lissajous orbit;
    \item The cycle or ellipse area size exactly similar to the Lissajous orbit;
    \item The ellipse major axis has some inclination angle with positive x-axis.
    \item The consequence of multi-frequency, the eigenvalue as $4i$ and $8i$,
    appears as (1:2)-Lissajous orbit.
    \end{enumerate}

    Notice that, this dynamics system do not satisfy the requirement of
    the central manifold requirement, because, one of the eigenvalue is
    positive. The (1:2)-Lissajous orbit pattern shown in figure
    \ref{fig:ode_invariant_manifold} is initial condition depended.
    On average, the zero hypnosis in Eq  \ref{eq:dele_noise}
    about Two components   cross contribution in delegation condition with noise.

\begin{footnotesize}
\begin{verbatim}
% Approximate invariant manifold for replicator dynamics equation.
% Matlab 2019b
function [T,Y] = tmp202106122()

tspan = [0 30];
ini = [2 1 1 1 2 1 1 1 ]/5 + 0.001*rand(1,8);
initial_pos = [ini(1:4)/sum(ini(1:4)) ini(5:8)/sum(ini(5:8))];

  [T,Y] = ode45(@osc,tspan,initial_pos);
        figure;plotmatrix(Y,'.-'); title(num2str(a))

        function dydt = osc(t,y)
dydt = ...
    [ y(1)*(5*y(5) - 5*y(6) - 5*y(7) - 5*y(8) + y(1)*(5*y(6) - 5*y(5) + 5*y(7) + 5*y(8)) + ...
      y(2)*(5*y(5) + 5*y(6) - 5*y(7) - 5*y(8)) + y(3)*(5*y(5) - 5*y(6) + 5*y(7) - 5*y(8)) + ...
      y(4)*(5*y(5) - 5*y(6) - 5*y(7) + 5*y(8)))
      y(2)*(5*y(7) - 5*y(6) - 5*y(5) + 5*y(8) + y(1)*(5*y(6) - 5*y(5) + 5*y(7) + 5*y(8)) +  ...
      y(2)*(5*y(5) + 5*y(6) - 5*y(7) - 5*y(8)) + y(3)*(5*y(5) - 5*y(6) + 5*y(7) - 5*y(8)) +  ...
      y(4)*(5*y(5) - 5*y(6) - 5*y(7) + 5*y(8)))
      y(3)*(5*y(6) - 5*y(5) - 5*y(7) + 5*y(8) + y(1)*(5*y(6) - 5*y(5) + 5*y(7) + 5*y(8)) +  ...
      y(2)*(5*y(5) + 5*y(6) - 5*y(7) - 5*y(8)) + y(3)*(5*y(5) - 5*y(6) + 5*y(7) - 5*y(8)) +  ...
      y(4)*(5*y(5) - 5*y(6) - 5*y(7) + 5*y(8)))
      y(4)*(5*y(6) - 5*y(5) + 5*y(7) - 5*y(8) + y(1)*(5*y(6) - 5*y(5) + 5*y(7) + 5*y(8)) +  ...
      y(2)*(5*y(5) + 5*y(6) - 5*y(7) - 5*y(8)) + y(3)*(5*y(5) - 5*y(6) + 5*y(7) - 5*y(8)) +  ...
      y(4)*(5*y(5) - 5*y(6) - 5*y(7) + 5*y(8)))
     -y(5)*(5*y(1) - 5*y(2) - 5*y(3) - 5*y(4) + y(5)*(5*y(2) - 5*y(1) + 5*y(3) + 5*y(4)) +  ...
      y(6)*(5*y(1) + 5*y(2) - 5*y(3) - 5*y(4)) + y(7)*(5*y(1) - 5*y(2) + 5*y(3) - 5*y(4)) +  ...
      y(8)*(5*y(1) - 5*y(2) - 5*y(3) + 5*y(4)))
     -y(6)*(5*y(3) - 5*y(2) - 5*y(1) + 5*y(4) + y(5)*(5*y(2) - 5*y(1) + 5*y(3) + 5*y(4)) +  ...
      y(6)*(5*y(1) + 5*y(2) - 5*y(3) - 5*y(4)) + y(7)*(5*y(1) - 5*y(2) + 5*y(3) - 5*y(4)) +  ...
      y(8)*(5*y(1) - 5*y(2) - 5*y(3) + 5*y(4)))
     -y(7)*(5*y(2) - 5*y(1) - 5*y(3) + 5*y(4) + y(5)*(5*y(2) - 5*y(1) + 5*y(3) + 5*y(4)) +  ...
      y(6)*(5*y(1) + 5*y(2) - 5*y(3) - 5*y(4)) + y(7)*(5*y(1) - 5*y(2) + 5*y(3) - 5*y(4)) +  ...
      y(8)*(5*y(1) - 5*y(2) - 5*y(3) + 5*y(4)))
     -y(8)*(5*y(2) - 5*y(1) + 5*y(3) - 5*y(4) + y(5)*(5*y(2) - 5*y(1) + 5*y(3) + 5*y(4)) +  ...
      y(6)*(5*y(1) + 5*y(2) - 5*y(3) - 5*y(4)) + y(7)*(5*y(1) - 5*y(2) + 5*y(3) - 5*y(4)) +  ...
      y(8)*(5*y(1) - 5*y(2) - 5*y(3) + 5*y(4)))
     ];
        end
end
\end{verbatim}
\end{footnotesize}

 \subsection{Appendix 7: Verification of the net transit and  eigencycle}\label{app:net_trans}

\begin{figure}[h!]
\centering
\includegraphics[scale=0.6,angle=90]{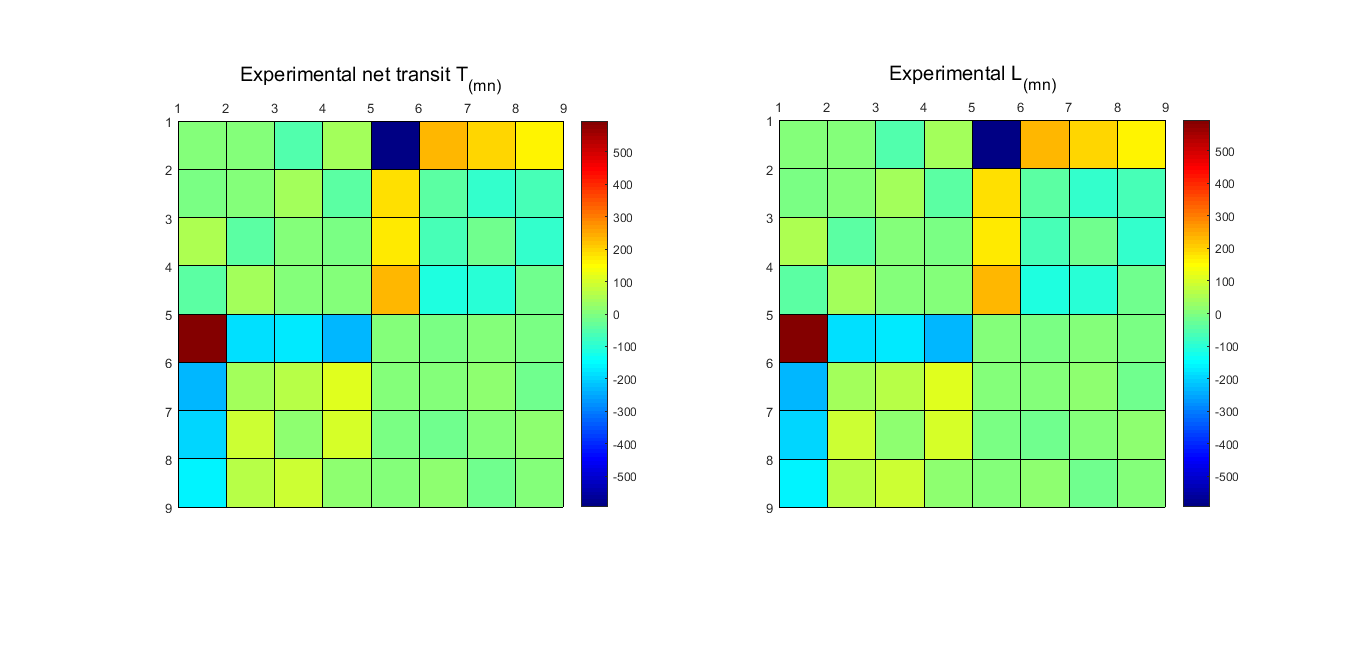}
\caption{Verification of the net transit and  eigencycle matrix.
The horizon-axis indicates $m$-dimension from left to right is 1 to 8,
and vertical-axis indicates $n$-dimension from up to down is 1 to 8.}
\label{fig:Tmn_Lmn}
\end{figure}

\begin{itemize}
\item \textbf{Definition}\\
Net transit (denoted as $T$) relates to the strategy probability transition,
which is a concept in stochastic processes.
The strategy probability transition from state $m$ to $n$,
which can be denoted as a Markovian matrix element $A_{m,n}$.
Denotes the stationary distribution as vector $\rho$,
the Net transit can be presented by the net transit matrix and defined as \cite{2010wangchinese}
$$ T_{m,n} = \rho_m A_{m,n} -  \rho_n A_{m,n}^T = \rho_m A_{m,n} -  \rho_n A_{n,m}, $$
which describes the net probability current $T$ from state-$m$ to state-$n$.

Assume a system has $N$ independent states total, this definition has following property.
\begin{itemize}
\item If the system is in detail balance condition,
        $$T_{m,n} =0, ~~~~ \forall {m,n} \in N; $$
\item If observed $T_{m,n} > 0$, there exists net current from $m$ to $n$;
\item If observed $T_{m,n} < 0$, there exists net current from $n$ to $m$;
\item The net transit matrix is anti-symmetry, $T_{m,n} = - T_{n,m}$.
\end{itemize}

Using the data shown in main text Table \ref{tab:datasource},
we can conduct following two calculation and present in matrix plotting,
\begin{itemize}
\item  The net transit matrix $T_{mn}$,
\item  The experimental angular momentum   $L_{mn}$.
\end{itemize}

\item \textbf{Result}\\
The results are shown in figure \ref{fig:Tmn_Lmn}.
In this plot,
we use all of the times series, total about 13 thousands transition.
The net transit appears in $T_{5,1}$ and $L_{5,1}$,
which means the probability of
the net transit counted is 500 in about 13 thousands transit.

Importantly, the comparison shows us the physical meaning
of the angular momentum,
and then the eigencycle, both of the value and the signal.

\end{itemize}

 \subsection{Appendix 5: Data and dynamics model valid}
 \begin{itemize}
 \item Data valid\\
        The two-person fixed-paired discrete time game experiment for testing the dynamical pattern.
        The literature \cite{wang2012127} shows that testing the dynamical pattern in a long-run fixed-paired 2 $\times$ 2 game in a strategy space is possible. Given the myopic nature of human behavior, we expect a cyclical pattern.
        Although the time series is highly stochastic, we can test dynamics behavior using the time reserve asymmetry measurement, much like the velocity field or angular momentum measurement \cite{WANG2017455}.
        In a one population symmetry 4$times$4 game experiments \cite{2021Shujie}, and in a one population symmetry 5$\times$5 game experiments \cite{2021Qinmei}, we have obtained that, eigencycle approach valid.
\item  Dynamics valid \\
        Employing replicator dynamics and perturbation expansion to the model.
         Replicator dynamics is the earliest developed model.
         Evidence shows that the model, as well as the perturbation expansion, can capture the main character of dynamics behavior \cite{wang2014}.
         Hence, both replicator dynamics and perturbation expansion can be part of standard analysis in mathematics, thereby offering a clear and practical picture of the eigen system.
 \end{itemize}

\newpage
\subsection{Appendix 3: Related works}
\subsubsection{Related works on the evolutionary game theory and experiment}
    ~\\
    \textbf{[Question]}
Main result of this paper is finding the dynamics cycle in high dimension strategy interaction system (game).
Interestingly, in quite recently paper \cite{dan2020}, the authors clearly pointed out that detecting cycles in high dimensions game remains an open question. This is the first question.

The second question is a basic in the field of evolutionary game theory and experiment.
The question is excellently abstracted by Dan Friedman and his colleague as \cite{dan2020}
\begin{center}
\textit{In the background is a metaphysical question: what is a cycle?}
\end{center}

~\\
\textbf{[History]}
In fact, the questions have last for nearly three decades since early 1990s when
evolutionary game theory came into economics \cite{dan1991,dan1998}. The time line on the questions can be divided into periods.

\begin{itemize}
    \item \textbf{No cycle observed period} is during 1990s-2000s, the common knowledge was, like mentioned
    in the textbook namely behavior game experiments \cite{Behavioral2003},
    evolutionary approach fit biology, or culture evolution, not fit to describe
    quickly human game strategy interaction behavior.
    Such description is based on the previous experiments reports.
    \item \textbf{Confirm cycle existence period} since 2010s, which mainly base on rock-paper-scissors game or 2 $\times$ 2 game experiment,
    no high dimension cycles reports.
        \begin{itemize}
            \item \cite{2015nowak} reported observation of distribution differs from Nash equilibrium concept but fit evolutionary dynamics in RPS game experiment.
            \item \cite{dan2014} reported observation of cycles in the continuous time setting  RPS game experiment;
            \item \cite{wang2014} reported observation of cycles in classical discrete time setting RPS game experiment;
            \item \cite{wang2014} reported observation of cycles in classical discrete time randomly matching 2 $\times$ 2 game experiment;
        \end{itemize}
    \item \textbf{Quantifying cycle measurement}, whenever report cycle existence,
           is necessary.
           Following measurements have been shown
           \begin{itemize}
               \item CRI (cycle rotation index) in RPS \cite{dan2014}
               \item period of cycle (or frequency of cycle) in
               2 $\times$ 2 \cite{wang2014}.
               \item velocity field of cycle in RPS \cite{WANG2017455}
               \item angular momentum of cycle in RPS \cite{WANG2017455}
               \item net transit between states in 2 $\times$ 2 game \cite{2010wangchinese}.
           \end{itemize}
    \item \textbf{Unifying cycle measurement}  has been concerned naturally. The unify between velocity and angular momentum \cite{WANG2017455},
    between CRI and frequency of cycle \cite{wang2014},
    has shown us the inner consistence of these measurements.
\end{itemize}
    ~\\
    \textbf{[Summery]} \\
    \begin{itemize}
        \item   On the open question about how to identify high dimension game, we hope, our finding of the eigencycle is an appreciate answer.
        \item  On the metaphysical question about what is a cycle,
        we suggest, with more accuracy results of more 'unexpected' observations from  deeper theoretical analysis and more experiments conduction,
        as the full picture being clear,
        the answer will become clearer and clear.
    \end{itemize}

\subsubsection{Related works on the dynamics system theory}\label{app:rw_dst}

Our study is not lonely on using the concepts and tools
from the dynamics system theory to study human  game dynamics behavior
in laboratory experiments since 1990s. The framework and the method
is common, which has been the standard narration mode \cite{dan2016book,2011Sandholm}.

 ~\\
\textbf{The framework} to describe the dynamics includes two concepts,
\begin{enumerate}
\item Game dynamics study has its own conventions since its very beginning,  by describing the dynamical systems with ODE (ordinary differential equations) mathematically;
\item The phase space of a dynamical system is the collection of all possible world-states of the system in question. Each world state represents a complete snapshot of the system at some moment in time;
\end{enumerate}
\textbf{The technology methods}, as well as its  illustrations in existed literature, are:
\begin{enumerate}
\item solve the equations for the rest points (or singularity point, equilibrium, stationary state);
\item analysis the evolutionary trajectory in the phase space;
\item solve the character matrix (Jacobian) for eigen system
      including eigenvalues and eigenvectors;
\item by real part of the eigenvalue, analysis the stability ;
\item by imagine part of the eigenvalue,
      evaluate the periods of limit cycle near the rest point;
\item linear decomposition the system by the eigen system, namely spectrum analysis.
\end{enumerate}

To our knowledge, none of the literature, at least in game dynamics theory and experiment fields, has focus on the eigenvector structure.

~\\
\textbf{Eigen system as analysis framework and toolbox}
Basing on linear approximation decomposition approach,
eigenvalue and eigenvector are
the corner stones for the formal dynamics system analysis framework.
The eigencycle approach, which is based on the eigenvector components, belongs to this framework.
Following, we revisit related approach in the framework.

\begin{itemize}
    \item Eigenvalue used as an analysis tool is common
    for spectrum analysis and stability analysis.
    As well known that, its real part determines the stability,
    meanwhile, its imaginary part determine the frequency of
    the vibration mode.
    \item Using real eigenvector is common too,
    for example, eigenvectors indicates distribution \cite{dan2016book,2011Sandholm}. The dynamic decomposition weights (DDW) approach \cite{2020system}
      using only the real part (not the imagine part) of the components of an eigenvector.
    \item Using the complex  characters of an eigenvector component is rare. Two welcome exceptions are (1)
 It has noticed the coupling of the eigenvalues to eigenvectors will lead to phase shift \cite{2009gonccalves}, but ignore the phase interfere between two eigenvector components.
 (2) Scholars have suggest to borrow quantum physics concepts on eigenvector to investigate \cite{2020epl}, but do not provide practical way or   example.
\end{itemize}
Comparing with existed approach,
our eigencycle approach is using
the phase interfere between the eigenvector components.
The potential power of the eigencycle approach is encouraged
by our surprisingly finding the high dimensional fine dynamics structure
our of decades long fundamental human subject game experiments.

~\\
\textbf{Advantage of the eigencycle}
\begin{itemize}
\item Eigencycle can be more efficient bridge between theory and experiment. In this study, we have illustrated the consistence of
the eigencycle with following two existed measurement,
\begin{enumerate}
\item the angular momentum measurement $L$, details are
shown in section  \ref{app:im_ec}.
\item the net transit measurement $T$, details are
shown in section \ref{app:net_trans}
\end{enumerate}
But different from these two measurements, the eigencycle set can be deviate from the eigenvectors, which can be explicitly calculate out of the dynamics equation set.
\item  Eigencycle can be an efficient tube to bring in the concepts and tools,
      for game dynamics research.
      The concepts and tools has already in the dynamics system theory,
      like central manifold theorem, invariant set theorem, eigen space,
      the invariant manifold and among others, we believe,
      will significantly improve our design and analysis
      experiment and theory .

\end{itemize}

\subsubsection{Related works on the dynamics cycle in O'Neill game}
The eigencycle approach bases on the eigenvector,  which rooting in dynamical system theory. We review the literature history in O'Neill game analysis to explain the
advance made by the eigencycle approach. As illustrated in figure \ref{fig:improve},  an extending explanation are follow.

\begin{itemize}
        \item Subfigure (1987) --- The above game matrix designed by O'Neill is historic. It provides the first evidence on the reality and accuracy of the mixed strategy Nash equilibrium. Subfigure (1978) ---
        The evolutionary character, when transited to a 2 $\times$ 2 game, can be captured by the replicator dynamics equation, as first noted in 1978 \cite{2011Sandholm}.
        \item  Subfigure (2001) --- This figure is derived from \cite{Binmore2001Minimax}. Regarding geometric cyclic behavior, where it was first presented, the authors conducted an experiment of 150 rounds, repeated per session, with a total of 13 sessions. However, their measurement does not yield results with significant confidence from the 13 sessions. Instead, they report manually selected 150 rounds with likely-cycle trajectory only.
       \item  Subfigure (2012) --- Based on the method to concentrate the 4$\times$4 zero sum game to a  2$\times$2 zero sum game, Wang and Xu report the existence of a cycle in the O'Neill game by testing the accumulated angular momentum \cite{2012arXiv1203.2591W}. As shown in Subfigure (2012), for all the 13 sessions, the accumulated angular momentum is constantly declining. That is, clockwise rotation is robust for all 13 sessions. However, as the second, third, and fourth strategies are concentrated in one strategy, the authors miss the high-dimensional cyclical characters. To elaborate, as all the
        second, third, and fourth strategies of player A and player B are combined into one strategy, namely, the down-strategy for A and down-strategy for B, the cyclic behavior in the combined strategies is missed.
     \item Subfigure (2020) --- By applying the eigencycle set analysis for the O'Neill game, there is no need to resort to the combination strategy of 2-3-4 and 6-7-8. It is possible to test all the cycles in all the 28 two-dimensional subspaces. We can also archive the discovery of the fine stricture.
\end{itemize}
\begin{figure}[h!]
\centering
\includegraphics[scale=0.65]{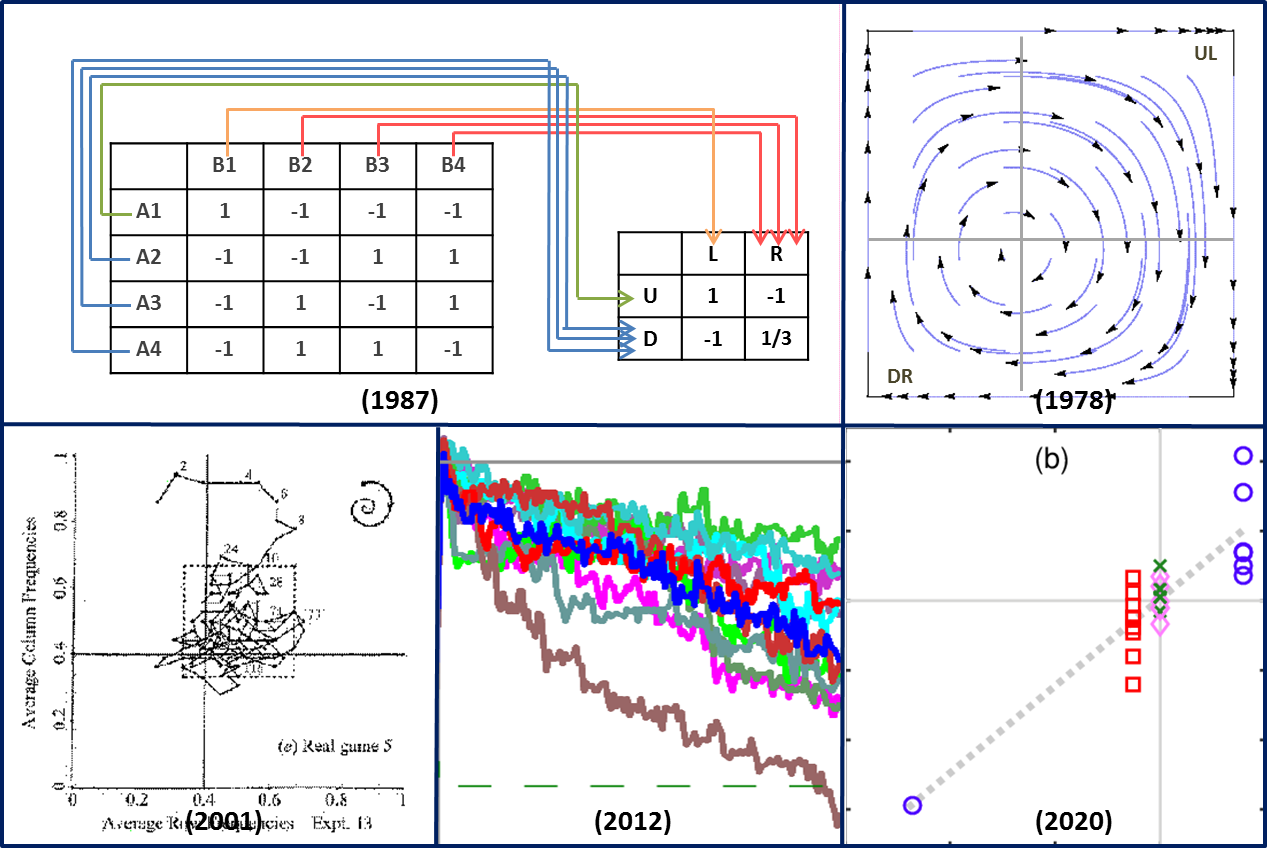}
\caption{(1987) The O'Neill 4$\times$4 original matrix and Binmore 2$\times$2 concentration matrix.
(1978) The replicator dynamics trajectory by Binmore's 2$\times$2 concentration matrix.
(2001) The figure (from \cite{Binmore2001Minimax}) with two-rounds average trajectory from a manual selection session, with unclear existence of a cycle in the long run.
(2012) The figure (from \cite{2012arXiv1203.2591W}) reports the 150-round accumulated angular momentum of the 13 sessions in the two-dimensional presentation, with the accumulated angular momentum in abscissa and the experiment time (rounds)
in  ordinate dimension.
(2020) The result from the eigencycle while exploring the eight-dimensional decomposed cycles, with theoretical eigencycle value $\sigma$ in abscissa and experimental angular momentum $L$ in  ordinate dimension.
}
\label{fig:improve}
\end{figure}

\subsubsection{Related work on high dimension game dynamics pattern}

\begin{itemize}
  \item  In natural science, there are many real examples.
In ecology, game dynamics pattern, as rock-paper-scissors,
 has been seen in male lizard competing \cite{dan2016book}.
If a ecology have more spices take into account,
higher dimension dynamics analysis technologies are needed.
  \item
In economics, there are many practical examples, too \cite{dan2016book}..
In a financial markets system, or supply chain, or public goods providing and protecting,
there are always many players and each player has many strategy.
To describe such systems, high dimensional dynamics analysis technologies are needed too.
  \item
In artificial intelligence study, this is also visible
If the game is approximately transitive,
then self-play generates sequences of agents of increasing strength.
However, nontransitive games, such as rock-paper-scissors in
poker or StarCraft \cite{2019deepmind}\cite{2020Grandmaster}\cite{2018poker}.
can exhibit a dynamics structure (strategic cycles),
 and there is no longer a clear objective 
 and against whom is unclear. 
 In such game system, the dimension is extremely high, 
 to test out the nontransitive strategy (component) is cruelly important.
\end{itemize}

\newpage

\end{document}